\documentstyle[12pt,epsfig]{ioplppt}

\newcommand {\om}  {\omega}
\newcommand {\Om}  {\Omega}
\newcommand {\calN}  {{\cal N}}
\newcommand {\E}  {\varepsilon}
\newcommand {\coul}  {\Lambda}
\newcommand {\ch}  {{\rm ch}}
\newcommand {\dUmax} {{U_{\max}^{\prime}}}
\newcommand {\Ld} {{L_{\rm d}}}

\begin{document}

\sloppy

\jl{2}

\title{Channeling of Positrons through Periodically Bent Crystals:
on Feasibility of Crystalline Undulator and Gamma-Laser
} 
\author{
A. V. Korol\dag\ftnote{4}
{E-mail: korol@rpro.ioffe.rssi.ru},
A. V. Solov'yov\ddag\S\ftnote{5}
{E-mail: solovyov@rpro.ioffe.rssi.ru} 
and
W.Greiner\S
}

\address{\dag Department of Physics,
St.Petersburg State Maritime Technical University,
Leninskii prospect 101, St. Petersburg 198262, Russia}

\address{\ddag A.F.Ioffe Physical-Technical Institute of the Academy
of Sciences of Russia, Polytechnicheskaya 26, St. Petersburg 194021,
Russia}

\address{\S Institut f\"ur Theoretische Physik der Johann Wolfgang
Goethe-Universit\"at,
60054 Frankfurt am Main, Germany}
 
\begin{abstract}
The electromagnetic radiation generated by ultra-relativistic 
positrons channelling in a crystalline undulator is discussed. 
The crystalline undulator is a crystal whose planes are 
bent periodically with the amplitude much larger than the 
interplanar spacing. 
Various conditions and criteria to be fulfilled for
the crystalline undulator operation are established.
Different methods of the crystal bending are described. 
We present the results of numeric calculations of spectral distributions 
of the spontaneous radiation emitted in the crystalline undulator
and discuss the possibility to create the stimulated
emission in such a system in analogy with the free electron laser. 
A careful literature survey covering the formulation of all 
essential ideas in this field is given. 
Our investigation shows that the proposed 
mechanism provides an efficient source for high energy photons, 
which is worth to study experimentally.
\end{abstract}

\section{Introduction}\label{Introduction}

We discuss a new mechanism of generation of high energy photons by
means of a planar channeling of ultra-relativistic positrons through a
periodically bent crystal.  
The {\it feasibility} of this scheme was explicitly demonstrated in 
\cite{Korol98,Korol99}.  
In these papers as well as in our subsequent publications
\cite{EnLoss00}-\cite{KorolSolovyovGreiner2003_2} the idea of this new type of
radiation, all essential conditions and limitations which must be
fulfilled to make possible the observation of the effect and a
crystalline undulator operation were formulated in a complete and
adequate form for the first time. 
A number of corresponding novel numerical results were presented to 
illustrate the developed theory, including, in particular, the 
calculation of the spectral and angular characteristics of the new 
type of radiation.

The aim of this paper is to review the results obtained so far in this
newly arisen field as well as to carry out a historical survey of the
development of all principal ideas and related phenomena.  
The necessity of this is motivated by the fact that the importance of the
ideas suggested and discussed in 
\cite{Korol98}-\cite{KorolSolovyovGreiner2003_2} has
been also realized by other authors resulting in a significant
increase of the number of publications in the field within the last
3-5 years
\cite{AvakianGevorgyanIspirianIspirian1998}-\cite{KorhmazyanKorhmazyanBabadjanyan2004} 
but, unfortunately, often
without proper citation 
\cite{AvakianGevorgianIspirianIspirian2001}-\cite{KorhmazyanKorhmazyanBabadjanyan2004}.  
We review all
publications known to us which are relevant to the subject of our
research.

The main phenomenon to be discussed in this review is the radiation 
formed in a  crystalline undulator.
The term 'crystalline undulator' (introduced but not at all elaborated 
clearly in \cite{KaplinPlotnikovVorobev1980}) stands for a system which 
consists of two essential parts: (a) a crystal whose crystallographic 
planes are bent periodically, and (b) a bunch of ultra-relativistic 
positively  charged particles undergoing planar channeling in the crystal.
In such a system there appears, in addition to a well-known channeling 
radiation, the radiation of an undulator type which is due to the 
periodic motion of channeling particles which follow the bending of 
the crystallographic planes. 
The intensity and the characteristic frequencies of this undulator 
radiation can be easily varied by changing the energy of beam particles
and the parameters of crystal bending.
\begin{figure}
\begin{center}
\epsfig{file=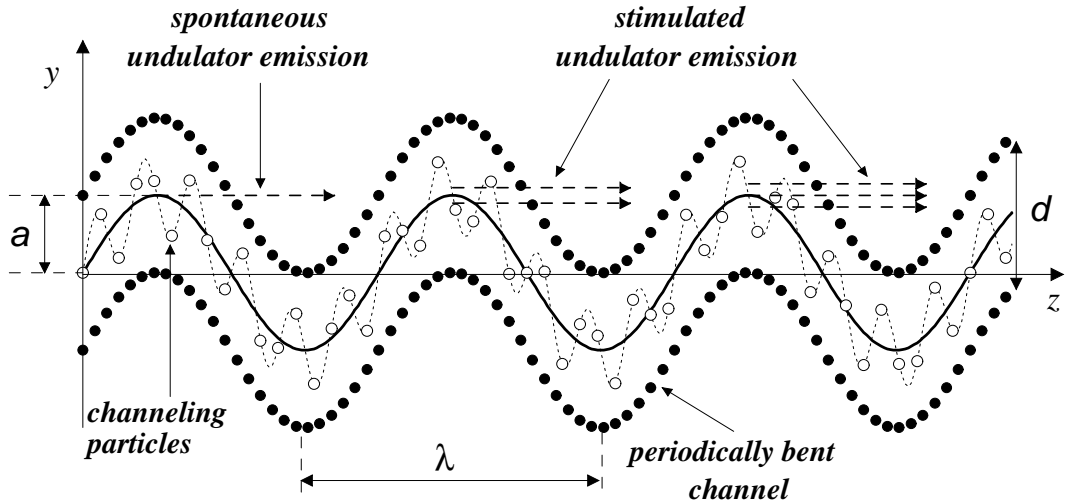,width=14cm}
\end{center}
\caption{Schematic representation of spontaneous and stimulated
radiation in a periodically bent
crystal. The $y$- and $z$-scales are incompatible!}
\label{figure1_jpg}
\end{figure}

The mechanism of the photon emission by means of a crystalline
undulator is illustrated by figure \ref{figure1_jpg}.  
Short comments presented below aim to focus on the principal 
features  of the scheme relevant to the subject of the review. 
At the moment we do not elaborate all the important 
details but do this in section \ref{Feasibility}.

The $(yz)$-plane in the figure is a cross section of 
an initially linear crystal,
and the $z$-axis represents the cross section of a midplane of two
neighbouring non-deformed crystallographic planes (not drawn in 
the figure) spaced by the interplanar distance $d$.
Two sets of black circles denote the nuclei which belong to the
periodically bent neighbouring planes which form a periodically bent 
channel. 
The amplitude of the bending, $a$, is defined as a maximum displacement
of the deformed midplane (thick solid line) from the $z$-axis.
The quantity $\lambda$ stands for a spatial period of the bending.
In principle, it is possible to consider various shapes, $y(z)$, of the 
periodically bent midplane.
In this review we will mainly discuss the harmonic form of this function,
$y(z) = a\sin(2\pi z/\lambda)$.
For further referencing let us stress here that 
we will mainly consider the case when the quantities  $d$, $a$ and $\lambda$ 
satisfy strong double inequality: $d \ll a \ll \lambda$. 
Typically $d\sim 10^{-8}$ cm, $a \sim 10\dots 10^2\ d$, 
and $a \sim 10^{-5} \dots 10^{-4}\, \lambda$.

Open circles in figure \ref{figure1_jpg} denote the channeled 
ultra-relativistic particles.
Provided certain conditions are met, 
the particles, injected into the crystal,
will undergo channeling in the periodically bent channel.
Thus, the trajectory of a particle (represented schematically by the 
dashed line) contains two elements.
Firstly, there are oscillations inside the channel due to the action of
the interplanar potential,\ - \ the channeling  oscillations.
This mode is characterized by a frequency $\Om_{\ch}$ dependent on 
the projectile type, energy, and the parameters of the interplanar potential.
Secondly, there are oscillations because of the
periodicity of the distorted midplane, - the undulator oscillations, whose
frequency is $\om_0\approx2\pi c/\lambda$
($c$ is the velocity of light which approximately is the velocity of  an 
ultra-relativistic particle).

Spontaneous emission of photons which appears in this system is 
associated  with both of these oscillations. 
Typical frequency of the emission due to the channeling oscillations 
is $\om_{\ch}\approx 2\gamma^2\Om_{\ch}$ where $\gamma$ is the relativistic
Lorenz factor $\gamma=\E/mc^2$.
The undulator oscillations give rise to the photons with  
frequency $\om \approx 4\gamma^2\om_0/(2+p^2)$ where the quantity
$p$, a so-called undulator parameter, is related to the amplitude and the
period of bending,  $p=2\pi\gamma (a/\lambda)$.

If strong inequality $\om_0 \ll \Om_{\ch}$ is met than the 
frequencies of the channeling radiation and the undulator radiation are 
also well separated, $\om \ll \om_{\ch}$.
In this case the characteristics of the undulator radiation 
(the intensity and spectral-angular distribution) are practically 
independent on the channeling oscillations but depend on the shape 
of the periodically bent midplane.

For $\om_0\ll \Om_{\ch}$ the scheme presented in figure
\ref{figure1_jpg} leads to the possibility 
of generating a stimulated undulator emission. 
This is due to the fact, that photons emitted at
the points of the maximum curvature of the midplane travel almost
parallel to the beam and thus, stimulate the photon generation in the
vicinity of all successive maxima and minima.
In \cite{Korol99} we demonstrated that it is feasible to consider
emission stimulation within the range of photon energies
$\hbar \om = 10\dots 10^4$ keV (a Gamma-laser).
These energies correspond to the range $10^{-8}\dots10^{-4}$ $\mu m$ 
of the emission wavelength, which is far below than the 
operating wavelengths in conventional free-electron lasers 
(both existing and proposed) based on the action of a magnetic field on 
a projectile \cite{TESLA,RullhusenArtruDhez,ColsonFELs}.

However, there are essential features which distinguish a seemingly 
simple scheme presented in figure \ref{figure1_jpg} from 
a conventional  undulator.
In the latter the beam of particles and the photon flux move in vacuum 
whereas in the proposed scheme they propagate through a crystalline media.  
The interaction of both beams with the crystal constituents makes the 
problem much more complicated from theoretical, experimental and technical
viewpoints.  
Taking into consideration a number of side effects which
accompany the beams dynamics, it is not at all evident 
{\it a priori} that the effect will not be smeared out.   
Therefore, to prove that the crystalline undulator as well as 
the radiation formed in it both are feasible it is necessary to analyze 
the influence, in most cases destructive, of various related phenomena.
Only on the  basis of such an analysis one can formulate the 
conditions which must be met and define the ranges of parameters 
(which include the bunch energy, the types
of projectiles, the amplitude and the period of bendings, 
the crystal length, the photon energy) within which all the criteria
are fulfilled. 
In full this accurate analysis was carried out very recently
and  the feasibility of the crystalline undulator and the Gamma-laser
based on it were demonstrated in an adequate form for the first 
time in \cite{Korol98,Korol99} and in 
\cite{EnLoss00}-\cite{KorolSolovyovGreiner2003_2}.
                                         
>From the viewpoint of this compulsory programme which had to be done in 
order to draw a conclusion that the scheme in 
figure  \ref{figure1_jpg} can be transformed from the stage of a 
purely academic idea up to an observable effect and an operating device
we review critically some of the recent publications 
as well as the much earlier ones
\cite{KaplinPlotnikovVorobev1980,BaryshevskyDubovskayaGrubich1980,
IkeziLinLiuOhkawa1984,BogaczKetterson1986,
Armyane1986,
AmatuniElbakyan1988,
MkrtchyanEtal1988,
Dedkov94}.
This is done in section \ref{Feasibility}.

Prior to start with the discussion of various
aspects of the electromagnetic radiation from a beam of charged
particles channeling in a periodically bent crystal 
let us briefly mention the phenomena closely related to
our main problem. 
These are: channeling in straight and in bent crystals, 
channeling  radiation and undulator radiation.

\section{Channeling, channeling radiation, undulator radiation}
\label{section:Channeling}

We do not pretend to cover the whole range of problems
concerning the channeling effect, the channeling radiation and
the radiation formed in the undulators.
This section is devoted solely for a brief description of the 
effects closely related to the main subject of this review.

\subsection{Channeling} \label{Channeling}

The basic effect of the {\it channeling process in a straight crystal}
is that a charged particle can penetrate at an anomalously large 
distance when traveling nearly parallel to a crystallographic
plane (the planar channeling) or an axis (the axial channeling) 
and experiencing the collective action of the electrostatic field of the 
lattice ions.
The latter is repulsive at small distances for positively charged 
particles and, therefore, they are steered into the interatomic region,
while negatively charged projectiles move in the close vicinity of ion
strings or planes.

Channeling  was discovered in the early 1960s by  computer
simulations of ion motion in crystals of various structure
\cite{RobinsonOen1963}.
Large penetration lengths were obtained for low-energy 
(up to 10 keV) ions incident along crystallographic directions of low 
indexes.
These calculations allowed to explain the results of earlier 
experiments.
Later, a comprehensive theoretical study \cite{Lindhard}
introduced the important model of continuum potential for the interaction
energetic charged projectiles and lattice atoms arranged in strings 
and planes. 
Using this approach the criterion for a stable channeling was formulated. 
According to it  the beam of particles becomes trapped by 
the interplanar or axial potential if the incident
angle of the beam with respect to the crystal plane or the axis is 
smaller than the so-called critical Lindhard angle $\theta_L$.
These concepts were subsequently widely used to interprete 
channeling experiments for low energy projectiles ($\E < 1$ GeV), 
see, e.g., \cite{Gemmell}, and
for high  energies of the particles ($\E=1\dots 10^2$ GeV)
\cite{RelCha,SorensenUggerhoj1989,Instrum,BaurichterBiinoEtal2000,
BiryukovChesnokovKotovBook}.
In recent years channeling of charged particles in straight nano-tubes, 
usually carbon nano-tubes, has been intensively investigated 
\cite{KlimovLetokhov}-\cite{ZhevagoGlebov2003}.
We mention these papers for the sake of completeness but 
do not pretend to review them, since this field is not related
directly to the main subject of our research.

Lindhard \cite{Lindhard} was also the first who theoretically 
described the {\it dechanneling process}, - the phenomenon of a gradual 
increase of the transverse energy of a channeled particle due to 
inelastic collisions with nuclei and electrons of the crystal.
As a consequence, initially channeled particle during its passage 
through the crystal gains the transverse energy 
higher than the continuum potential barrier. 
At this point the particle leaves the channel and is, basically, lost
for the channeling process. 
The scale which defines the (average) interval for a particle to penetrate
into a crystal until its dechannels is called a dechanneling length,
$\Ld$.
This quantity depends on the parameters of a crystal 
(these include the charge of nuclei, the interatomic spacing, 
the mean atomic radius, the amplitude of thermal vibrations) and 
 the parameters of a channeled particle, - the energy and the charge. 
It is important to note that for negative and for positive projectiles
the dechanneling occurs in different regimes.
As mentioned above, the interplanar (or axial) potential is 
repulsive at small distances for positively charged 
particles and is attractive for negatively ones. 
Therefore, negatively charged particles tend to channel in the regions 
around the nuclei whereas positive particles are pushed away.
Consequently, the number of collisions with the crystal constituents 
is much larger for negatively charge particles and they dechannel
faster.
Typically, the dechanneling lengths of positive charges exceed those
for negative ones (of the same energy and charge modulus) by the order 
of magnitude or more \cite{MikkelsenUggerhoj2000,Kumakhov2}.
This statement is valid, both for axial and planar channeling, and for 
various pairs of positive/negative particles of the same charge modulus:
for $e^{+}/e^{-}$ \cite{Uggerhoj1980}, 
 $\pi^{+}/\pi^{-}$ \cite{AndersenFichNielsenEtAl1980},
and $p/\bar{p}$ \cite{Uggerhoj2003}.

\subsection{Channeling radiation}
\label{ChannelingRadiation}

Channeling of charged particles in crystals
is accompanied by the {\it channeling radiation} \cite{Kumakhov1,Barysh77}.
This specific type of electromagnetic radiation arises due to the
transverse motion of the particle inside the channel under
the action of the interplanar field (the channeling oscillations, see
figure \ref{figure1_jpg}).
The phenomena of channeling radiation of a charged projectile in
a linear crystal, see e.g. 
\cite{SorensenUggerhoj1989,Instrum,Kumakhov2,
Baryshevsky_Kniga,Kniga,Baier,
Berman_Nato2002_review},
as well as in a 'simple' (i.e. non-periodic, one-arc) bent channel
\cite{TaratinVorobiev,Arutyunov,SolovyovSchaferGreiner}, are known, 
although in the latter case the theoretical and experimental data are 
scarce, at least up to now.

For the purpose of this review it is important to mention several 
well-established features of the channeling radiation.
Firstly, as well as in any other radiative process of a charge 
moving in an external potential (e.g. bremsstrahlung \cite{Land4}),
the intensity of the channeling radiation is inversely proportional
to the squared mass of the projectile.
Consequently, a channeled electron/positron emits $(m_p/m_e)^2$ times more
intensively than a proton with the same value of relativistic factor 
$\gamma$. 
Secondly, for both electrons and positrons the intensity of radiation in
the channeling mode greatly exceeds (by more than an order of magnitude)
that by the same projectile in an amorphous medium
\cite{Uggerhoj1980,BakEtal1985}.
Finally, the radiative energy loss is of a channeled
electron is noticeably higher than that of a positron of the 
same energy.
This is valid for all energy ranges (from several MeV up to hundreds of 
GeV) of the projectile 
\cite{MikkelsenUggerhoj2000,SorensenUggerhoj1989,BakEtal1985,Andersen1980,
ChanRadEkspPhysLett1991,KirsebomMikkelsenUggerhoj2001}.
Physical reason for this is that electrons channel in vicinity of
the crystal nuclei, therefore they are accelerated stronger and radiated 
more energetic photons than positrons which channel
in the region of a weaker field.

The study of the channeling radiation initially proposed for 
particles moving in crystals was later extended to the 
case of nanotubes 
\cite{KlimovLetokhov,Gevorgyan97-98,ZhevagoGlebov,Dedkov1998}.

\subsection{Channeling in bent crystals}
\label{ChannInBentCrystals}

The channeling process in a bent crystal takes place \cite{Tsyganov} 
if the maximal centrifugal force, acting on the projectile because of the
channel bending, is less than the force due to the interplanar field 
\cite{BaryshevskyDubovskayaGrubich1980,SolovyovSchaferGreiner,Tsyganov,
Kudo1981}:
\begin{equation}
m \gamma v^2/R < q\, U_{\rm max}^{\prime}.
\label{1}
\end{equation}
Here $m$ and $v$ are the mass and velocity of the projectile,
$\gamma$ is its relativistic factor and
$R$  is  a curvature radius of the bent channel,
$q$ is the charge of the projectile and the quantity
$U_{\rm max}^{\prime}$ stands for the maximum gradient of the
interplanar field.

Provided this condition is fulfilled, the beam of positively charged 
channeling particles at each instant moves inside the channel, especially 
parallel to the bent crystal midplane as it does in the case of 
the channeling in a straight crystal. 
Since it first theoretical prediction \cite{Tsyganov} and 
experimental support \cite{VodopyanovGolovatyukElishevEtal1979}
the idea to deflect high-energy beams of charged particles 
by means of a planar channeling in bent crystals 
had attracted a lot of attention worldwide 
and still is of a great interest. 
Indeed, the beams (in particular, those of ultra-relativistic 
protons and heavy ions) can be steered by crystals much more
efficiently than by means of external macroscopic electric or 
magnetic fields.
In recent experiments with 450 GeV protons 
\cite{MollerWormClementEtAl1994,Biino} the efficiency
of the particle beam deflection was reported on the level of 60\% 
\cite{Biino} and
of 85\% for the 70-GeV protons beam \cite{Afonin}.
The progress in deflecting of heavy-ions beams was reported recently as well
\cite{Uggerhoj}.  
We refer to the reviews papers which cover state-of-the-art in this field, 
and describe in detail all the stages of the development of the idea as well
as a number of phenomena which are of current interest
\cite{RelCha,Instrum,BaurichterBiinoEtal2000,BiryukovChesnokovKotovBook,
Taratin1998,RHIC2003,Uggerhoj2003_Fra}.

Deflection of the beams of negatively charged particles by planar channeling 
is strongly suppressed because of the increased role of the dechanneling 
\cite{BaurichterBiinoEtal2000,Uggerhoj_pc}.
The deflection of beams of particles, both positive and negative,
during axial channeling was simulated numerically 
\cite{GreenenkoShulga2001}.
The experimental study of this effect has demonstrated small efficiency
of extracting the beam particles \cite{Uggerhoj2003_Fra,Uggerhoj_pc}.

Another possible application of bent crystal concerns its possible
use to focus beams of ultra-relativistic heavy-ions
\cite{SchaferGreiner1991} or protons \cite{DenisovFedinGordeevaEtAl1992}.
To this end a crystal is needed in which the crystal axes are no
longer parallel, but are slanted more and more the farther away they
are from the axis of the beam.
Then the bending angle of the particles far away from the beam axis would
be largest and a general focusing effect will result.
Such a crystal can in principle be produced by varying the nickel to copper
(or Sb to Bi) ratio in a mixed crystal \cite{SchaferGreiner1991}. 
Similar idea was suggested also for producing periodically bent crystals, see
section \ref{Feasibility} for more detail.

Recently, it was suggested  to use bent nanotubes to steer beams of 
charged particle \cite{GreenenkoShulga2002,ZhevagoGlebov2003}.

\subsection{Undulator radiation}

The theory and also various practical implementations of the
undulator radiation, i.e. the radiation emitted by a charge moving in
spatially periodic static magnetic fields (a magnetic undulator), or
in a static macroscopic electric field (electrostatic undulator), or in 
a laser field (a laser-based undulator), etc. have long history
\cite{Ginz,Motz} and are well elaborated
\cite{RullhusenArtruDhez,Kniga,Baier,Alferov,Fedorov,Revista}.
The most important feature of the undulator radiation, which clearly 
distinguishes it from other types of electromagnetic radiation formed
by a charge moving in external fields, is in a
peculiar form of the spectral-angular distribution.
Namely, for each value of the emission angle $\theta$ (measured
with respect to the undulator axis) the spectral distribution
consists of a set of narrow, powerful and equally spaced peaks (harmonics).
The peak intensity is proportional to the square of 
total number of the undulator periods, $N^2$.
This factor reflects the constructive interference of radiation emitted
from each of the undulator periods and is typical for any system which
contains $N$ coherent emitters (e.g. \cite{Landau2}).
In an ideal undulator, i.e. the one in which a particle follows either
a sinusoidal periodic trajectory (a planar undulator) or a helical one
(a helical undulator) under the action of a non-dissipative
external force,  and for a fixed $\theta$ the widths of all peaks 
are the same and proportional to $1/N$. 
The coherence is lost if one integrates the spectral-angular distribution 
over the emission angle.
Indeed, although the spectral distribution as a function of photon 
frequency $\om$ contains maxima (which are noticeably widened as compared to
those in spectral-angular distribution) the intensity in which
is proportional to $N$.

These two features of the undulator radiation, the high intensity in the
peaks of spectral-angular distribution and the narrow widths,
are important for a successful operation of free-electron
lasers (FEL), the devices which transform the spontaneous undulator 
radiation into the stimulated one. 
Since the first discovery of the FEL operational principle \cite{Madey71}
both the theory and practical implementation of FELs have been well 
elaborated  (see, e.g. 
\cite{RullhusenArtruDhez,ColsonFELs,Baier,Fedorov,Revista,
SaldinSchneidmillerYurkov1995}).

In ideal conditions a bunch of particle undergoing 
channeling in a linear crystal can be considered as an undulator
(a natural undulator).
Indeed, due to the periodicity of the trajectory the characteristics of 
the channeling radiation are close to those of the undulator radiation
\cite{Kumakhov2}.
However, the characteristics of the radiation emitted in natural
undulator are masked in experiment 
(e.g. \cite{SorensenUggerhoj1989}) because of
the distribution of the beam particles in the transverse energy, and in the 
incident angle. 
Additionally, in the natural undulator the emission peaks become noticeably
wider due to a deviation of an interplanar potential from a 
harmonic one.
This leads to the dependence of the undulator period on the amplitude of 
the channeling oscillations which, in turn, defines the frequency of the
emitted radiation. 

\section{Channeling in a periodically bent crystal: 
feasibility of a crystalline undulator}
\label{Feasibility}

In this section we describe, more thoroughly, the crystalline undulator,
formulate the  conditions, which must be fulfilled for its 
operation, and present the detailed literature survey covering
the development of all essential aspects of this important idea.  
Also we review theoretical methods to be used for an adequate 
description of this phenomenon.
The idea of a gamma-laser based on the crystalline undulator is
discussed in section \ref{StimulatedEmission}. 

\subsection{The conditions to be met in a crystalline undulator}
\label{Conditions}

As it was noted in section \ref{Introduction} in a crystalline undulator
the beam of channeled particles and the emitted photons
propagate in a media.
Therefore, prior to drawing a conclusion that the scheme illustrated by 
figure \ref{figure1_jpg} is not of academic interest but can be made 
realistic and to represent a new type of 
an undulator, one has to understand to what extent general characteristics
of the undulator radiation (high intensity, high degree of monochromaticity
of the spectral-angular distribution) are influenced by the presence of a 
crystalline media.

To fulfill this programme and to establish the ranges of various 
parameters within which the operation of the crystalline undulator
is feasible we had analyzed the following basic problems:
     \begin{itemize}
         \item Planar or axial channeling?  Positive or negative particles?
               \cite{Korol98,Korol99}
         \item Condition for the stable channeling in a periodically bent 
               crystal 
               \cite{Korol98,Korol99}.
         \item Crystalline undulator preparation: static and 
               dynamic bending 
               \cite{Korol98,Korol99,Erevan01,Darmstadt01}.
         \item Large and small amplitude regimes 
               \cite{Korol99,EnLoss00,TotSpect00,Dechan01,Durham00}.
         \item Dechanneling and photon attenuation and 
               the length of a crystalline undulator 
               \cite{Korol99,Dechan01,Denton00,Durham00}.
         \item Energy losses and the shape of a crystalline undulator 
               \cite{EnLoss00,Erevan01,Darmstadt01}.
     \end{itemize}

\subsubsection{Planar channeling of positively charged particles.}
\label{PlanarPositrons}

Keeping in mind that negatively charged particles
are steered along crystallographic axes and planes much less
efficiently than positively charged ones and that the effect of axial
channeling in bent crystals has not been observed so far,
we focus our discussion on the planar channeling of positively 
charged particles and, in particular, on the channeling of 
light projectiles, positrons, in periodically bent crystals. 
This system is the most appropriate for the creation of
the crystalline undulator operating in the high energy photon regime.
Note, that from theoretical viewpoint the generalization 
of the treatment of the undulator radiation in crystalline undulators
to the case of axial channeling is straightforward.
However, small values of the dechanneling lengths of projectiles 
in this case make such a discussion purely academic and bring it far 
beyond any realistic experimental opportunities.

The principal difference in the behaviour of positively and negatively 
charged particles in a crystalline undulator
was realized for the first time in \cite{Korol98,Korol99}. 
This fact has determined, to a great extent, the main focuses
of our subsequent publications  
\cite{EnLoss00}-\cite{KorolSolovyovGreiner2003_2}.
In some of the earlier publications  on the subject 
\cite{KaplinPlotnikovVorobev1980,IkeziLinLiuOhkawa1984,
BogaczKetterson1986,Dedkov94}, 
the authors did not distinguish clearly the cases of 
electron and positron channeling
in a crystalline undulator and often discussed channeling of electrons
rather than positrons. 
They did not analyze the strong condition,
limiting the operation of a crystalline undulator,
which originates from the dechanneling of particles during their passage
through periodically bent crystals. 
Below we consider this important condition in more detail 
when discussing the dechanneling process.
Following our conclusions made in 
\cite{Korol98,Korol99,EnLoss00,TotSpect00,Dechan01},
the authors of \cite{AvakianGevorgyanIspirianIspirian1998} 
in their later publications 
\cite{AvakianGevorgianIspirianIspirian2001,AvakianAvetyanIspirianMelikyan2002} 
also began to focus on the channeling of positively charged particles in 
crystalline undulators.  
This remark concerns not only this particular issue, but rather 
most of the physics of the processes taking place in crystalline undulators, 
which was analyzed and clarified in \cite{Korol98,Korol99} and our 
subsequent publications.
For example, all physical conditions for the operation of a
crystalline undulator outlined on page 113 in 
\cite{AvakianGevorgianIspirianIspirian2001} are
taken from our earlier work \cite{Korol99} but without any reference. 

\subsubsection{Stable channeling in a periodically bent crystal.}
\label{StableChanneling}

The condition for channeling in a periodically bent crystal 
is subject to the general criterion (\ref{1}) for the
channeling process in a bent crystal \cite{Tsyganov} (see also
\cite{BaryshevskyDubovskayaGrubich1980,BogaczKetterson1986,
BiryukovChesnokovKotovBook,SolovyovSchaferGreiner,
BaryshevskyDubovskaya1991}), 
and can be fulfilled by a proper
choice of the projectile energy and the maximal curvature of the channel.
For the first time the limiting role of the channeling criterion on
the parameters of the crystalline
undulator  was elucidated in 
\cite{Korol98,Korol99,TotSpect00}.
In the earlier publications on the subject this analysis
was not carried out.

Similar to the case of an one-arc bent crystal,
a stable channeling of an ultra-relativistic
positively charged particle in a periodically bent crystal occurs
if the maximum  centrifugal force, 
$F_{\mathrm{cf}}\approx m \gamma c^2/R_{\min}$, 
is less than the maximal force due to the
interplanar field, $F_{\mathrm{int}} = qU_{\max}^{\prime}$
\cite{Korol98,Korol99,BaryshevskyDubovskayaGrubich1980,BogaczKetterson1986}.
More specifically \cite{Korol99,EnLoss00,TotSpect00}, 
the ratio $C=F_{\mathrm{cf}}/F_{\mathrm{int}}$ is better to keep 
smaller than 0.1. 
If otherwise the phase volume of the trajectories, corresponding to the
channeling mode,  becomes significantly reduced. 
The inequality $C \ll 1$ relates the energy of an ultra-relativistic
 particle, $\E=m \gamma c^2$, the parameters of the bending 
(these define the quantity $R_{\min}$), and the characteristics of the
planar potential.

Choosing a harmonic shape $y(z)=a\sin\Bigl(2\pi z/\lambda\Bigr)$ 
for the periodically bent channel (see figure \ref{figure1_jpg}), one
derives $R_{\min}={\lambda^2 / 4\pi^2\,a}$.
Thus, the decrease in  $R_{\rm min}$ and, consequently, the increase
in the maximum centrifugal acceleration of the particle in the channel 
is achieved by decreasing $\lambda$ and increasing $a$.
The criterion for a stable channeling implies the following
relationship between the parameters $a$, $\lambda$, $\E$ and 
$U_{\max}^{\prime}$:
\begin{eqnarray}
C \equiv
(2\pi)^2{\E \over q U_{\max}^{\prime}}
{a\over \lambda^2} \ll 1.
\label{Condition.1}
\end{eqnarray}

For the first time, the limiting role of the channeling criterion 
(\ref{Condition.1}) on the parameters of  the crystalline
undulator  was elucidated in \cite{Korol98,Korol99,TotSpect00}.

Provided (\ref{Condition.1}) is fulfilled, the projectile, 
incident at the angle smaller than the Lindhard angle $\theta_L$ 
with respect to the mid-plane, is trapped in the channel, 
see figure \ref{figure1_jpg}.
The passage of an ultra-relativistic particle
gives rise to the emission of photons due to the
curvature of the trajectory, which becomes periodic reflecting 
the periodicity of the channel midplane.
This radiation is enhanced because of the coherent emission from similar 
parts of the trajectory, and, as a result, it may 
dominate over the channeling radiation 
\cite{Korol98,Korol99,EnLoss00,TotSpect00}.

\subsubsection{Large and small amplitude regimes.}
\label{LargeAndSmallAmpl}

There are two essentially different regimes of the radiation formation 
in a periodically bent crystals.
These regimes are defined by the magnitude of the ratio $a/d$.

It was demonstrated in 
\cite{Korol98,Korol99,TotSpect00,BaryshevskyDubovskayaGrubich1980}
that the undulator radiation and the 
channeling radiation are well separated provided 
the condition $a\gg d$ is fulfilled, see figure \ref{figure1_jpg}.
Typically, spacing between the planes, which are characterized by
the low values of the Miller indices such as the (100), (110) and (111)
planes, lies within the range $0.6\dots2.5$ \AA\  
(see, e.g. \cite{Gemmell,BiryukovChesnokovKotovBook,Baier}).
Therefore, the condition is fulfilled for the bending amplitudes
$a \geq 10$ \AA.
A similar separation of the two
radiative mechanisms takes place in non-periodically bent crystals, 
where the curvature of the channel leads to an
additional synchrotron-type radiation by a channeling particle
\cite{Arutyunov,SolovyovSchaferGreiner}.
This component of the emission transforms into the undulator radiation
in the periodically bent channel.

In the limit $a\gg d$, which we call a large-amplitude regime,
the separation of the two radiative mechanisms occurs
because the frequency of the channeling oscillations, 
$\Om_{\ch}\sim c(qU_{\max}^{\prime}/d \E)^{1/2}$,
is much higher than the frequency, $\om_0$, 
of the transverse oscillations caused by the periodicity of the channel
$\om_0 = 2\pi c/\lambda$. 
Consequently, the characteristic frequencies of the channeling  
and the undulator radiation can be estimated as 
$\om_{\ch}\approx\gamma^2\Om_{\ch}\sim\gamma^2c(q\dUmax/d\E)^{1/2}$ and
$\om_{\rm u} \sim \gamma^2\om_0 = 2\pi c\gamma^2/\lambda $.
Then, the ratio $\om_{\rm u}^2/\om_{\ch}^2$ can be written in the 
following form 
\begin{equation}
{\om_{\rm u}^2 \over \om_{\ch}^2}
\sim
{(2\pi)^2 \over \lambda^2}
{d\, \E \over q\dUmax}
=
C\,{d\over a}
\ll 1\, .
\label{Cond4}
\end{equation}
This relation shows that if both conditions, $C\ll 1$ and $a \gg d$,
are fulfilled, then  the characteristic frequencies are well separated.
Moreover, if one is only interested in the spectral distribution of the
undulator radiation, one may disregard the channeling oscillations and 
assume that the projectile moves along the centerline of the 
bent channel  \cite{Korol98,Korol99}.
Additionally, as it will be demonstrated below in the paper, 
in the high-amplitude regime
the intensity of the undulator radiation is higher 
than that of the channeling radiation \cite{Korol99,EnLoss00,TotSpect00}.
Therefore, we may state that the limit $a\gg d$ is essential to call the
undulator-type radiation due to the periodic structure of the crystal bending
as a new phenomenon and to consider it as a new source of
the emission within the $X$- and $\gamma$-range.

For $a \sim 10 d$ and for $C \leq 0.1$ equation (\ref{Cond4}) yields 
$\om_{\rm u}/\om_{\ch}\sim 0.1$.
Therefore, the two spectra are well  separated and the 
crystalline undulator radiation, both spontaneous and stimulated,
can be treated independently from the ordinary channeling radiation.

The importance of the large-amplitude regime was noted
in \cite{BaryshevskyDubovskayaGrubich1980} and independently in 
\cite{Korol98,Korol99}. 
In the papers \cite{Korol98,Korol99,TotSpect00,Dechan01}
the detailed qualitative and quantitative analysis of this condition 
was carried out, and the calculation of realistic spectral and 
angular distribution of the undulator radiation were performed for the 
first time.
The first calculation of the total radiative spectra
(i.e. the spectra in which both ordinary channeling radiation and 
the crystalline undulator radiation are taken into account)
for  the $\E=500$ MeV positrons channeling in the silicon
along the (110) crystallographic planes for different $a/d$ ratios
has been carried out in \cite{TotSpect00}. 
These calculations have supported the qualitative arguments, 
formulated above,
on the possibility to separate radiations emitted via two different
mechanisms. 
Calculations, performed in \cite{Korol98,Korol99,TotSpect00,Dechan01},  
have also demonstrated that the intensity of the crystalline 
undulator radiation can be made much larger than that of the 
channeling radiation. 
We consider this example  in more detail in section \ref{CrystUndRad}.

In the large-amplitude regime, in addition to the condition 
(\ref{Condition.1}), it is very important to consider other 
conditions under which the periodically bent crystal may serve as a 
crystalline undulator.
These other conditions appear when one takes into account 
the destructive role of the dechanneling effect and the photon
attenuation.
The restrictive influence of these phenomena on the 
parameters of the crystalline undulator is discussed in detail in
section \ref{DechannnelingAttenuation}.
Here we want to note that a comprehensive quantitative
analysis of all the conditions, which must be met,
was performed in \cite{Korol98,Korol99,EnLoss00,TotSpect00,Dechan01}.
In contrast such a full and important discussion was omitted in the papers
\cite{KaplinPlotnikovVorobev1980,BaryshevskyDubovskayaGrubich1980,
IkeziLinLiuOhkawa1984,BogaczKetterson1986,Dedkov94,Baryshevsky_Kniga}.

Our analysis, carried out in \cite{Korol99,TotSpect00,Dechan01}, 
has shown, in particular, that the optimal range of the amplitude values,
where all the necessary conditions can be fulfilled,
is $a\sim (10\dots 100)\, d \approx (10^{-7}\dots 10^{-6})\, {\rm cm}$.
We want to point out that exactly this range of $a$ 
was discussed in later publications 
\cite{AvakianGevorgianIspirianIspirian2001,
AvakianAvetyanIspirianMelikyan2001_2,
AvakianAvetyanIspirianMelikyan2002,
AvakianAvetyanIspirianMelikyan2003,
BellucciEtal2003,BellucciEtal2003_archive,Bellucci2003,
KorhmazyanKorhmazyanBabadjanyan2004}
without proper citation of our results.
The values of $a$ mentioned above are much lower than the interval 
$a\sim (10^{-5}\dots 10^{-4})\, {\rm cm}$ considered in 
\cite{BaryshevskyDubovskayaGrubich1980}.
In section \ref{AllConditions} we demonstrate that the parameters, used
in \cite{BaryshevskyDubovskayaGrubich1980} to characterize the crystal bending, 
do not lead to the emission of undulator-type radiation.

In the papers \cite{Korol99,TotSpect00,Dechan01} we demonstrated 
that in the low-amplitude regime, when $a < d$, 
the intensity of the undulator radiation is smaller 
than that of the channeling radiation.
Moreover, in the limit $a \ll d$ the undulator radiation 
becomes less intensive than the background bremsstrahlung radiation.
Hence, it is highly questionable whether the crystalline undulator radiation
can be considered as a new phenomenon in the limit of low $a$.
However, the low-amplitude regime  allows to consider a 
a resonant coupling of two mechanisms of 
the photon emission, the channeling radiation and the undulator radiation. 
Indeed, for $a \ll d$ and by a proper choice of the parameter $C$ (see
 (\ref{Condition.1})) it is possible to make the frequency of the undulator
radiation $\om_{\rm u}$ comparable or equal to that of the channeling 
radiation $\om_{\ch}$.
If $\om_{\rm u} \sim \om_{\ch}$, then the intensity of the {\it channeling} 
radiation can be resonantly enhanced even in the case when the 
undulator radiation is much less intensive.
This very interesting phenomenon  was considered in 
a series of papers 
\cite{IkeziLinLiuOhkawa1984,BogaczKetterson1986,Armyane1986,
AmatuniElbakyan1988,MkrtchyanEtal1988,Dedkov94,GrigoryanEtal2001,
GrigoryanEtal2003}.
The first two from this list discussed the 
modification of the channeling radiation spectrum 
in the case of an electron and/or a positron channeling in a superlattice.
In the papers  
\cite{Armyane1986,AmatuniElbakyan1988,MkrtchyanEtal1988,Dedkov94,
GrigoryanEtal2001,GrigoryanEtal2003} the parametric resonant 
enhancement of the channeling radiation emitted by positrons in the presence
of either transverse or longitudinal supersonic field was considered.
The authors of the cited papers investigated neither the crystalline 
undulator nor the undulator radiation formed in it.
Therfore, we are not going to discuss further the results 
obtained in these papers,
because the subject of their research is absolutely different from 
the topic of this review.

\subsubsection{Crystalline undulator preparation: static and 
          dynamic bending}
\label{StaticAndDynamic}

The term 'undulator' implies that the number of periodic elements (i.e.
the number of undulator periods, $N$) is large. 
Only in this limit the radiation formed during the passage of a 
bunch of relativistic particles through a periodic system 
bears the features of an undulator radiation 
(narrow, well-separated peaks in spectral-angular distribution)
rather than those of a synchrotron radiation.
Hence, the following strong inequality, which 
entangles the period $\lambda$ and the length of a crystal
$L$ must be met in the crystalline undulator \cite{Korol98,Korol99}:
\begin{eqnarray}
N = {L \over \lambda} \gg 1.
\label{Condition.2}
\end{eqnarray}

The parameter $\lambda$, together with the amplitude $a$ and the energy
of the particle, define other quantities, which are called
the undulator frequency $\om_0$ and parameter $p$,
and which are commonly used when characterizing an undulator:
\begin{equation}
\omega_0 = 2\pi\, {c \over \lambda},
\qquad
p = 2\pi\,\gamma {a \over \lambda} .
\label{Om0andP}
\end{equation}
Let us note the proportionality of the parameter $p$ to
the relativistic factor $\gamma$.
This dependence is typical for a crystalline undulator, but 
is absent for the undulators (both helical and planar) 
based on the action of magnetic field.
In the latter case the undulator parameter is equal to  
$p_{\rm B}= q B \lambda_{\rm B} / 2 \pi mc^2$
(see e.g. \cite{Revista,Colson85}),
where $B$ is the amplitude value of the magnetic induction and
$\lambda_{\rm B}$ is the period of the magnetic field. 

In turn, the quantities $\omega_0$, $p$ and $N$ 
define the parameters of the spontaneous undulator radiation,
which are the characteristic frequencies  
$\omega_k=k\, \om_1$, $k=1,2,\dots$
(harmonics) and the natural width, $\Delta\om$, of the peaks.
In a perfect planar undulator, where the trajectory of the particle 
has a harmonic form $y(z)=a\sin\Bigl(2\pi z/\lambda\Bigr)$,
the frequency $\om_1$ of the fundamental harmonic 
and the width are given by (see, e.g. \cite{Revista}):
\begin{eqnarray}
\cases{
\om_1
=
{4\gamma^2\, \omega_0 \over 2 + 2\,\gamma^2 \vartheta^2 + p^2},
\\
{\Delta\om  \over \om_1} = {2 \over N}.
}
\label{harmonics.1}
\end{eqnarray}
Here $\vartheta$ is the emission angle measured with respect to 
the $z$-direction (see figure \ref{figure1_jpg}).
Note that the relative width, $\Delta\om/\om_1$, is inversely proportional
to $N$. 

>From the general theory of a planar undulator 
(see, e.g. \cite{Baier,Revista}) it is known, that the magnitude of
the undulator parameter, $p$, defines the number of the harmonics in
which the radiation is effectively emitted.
In the case $p<1$ the radiation is mainly emitted into the first
harmonic. 
In the limit $p \gg 1$ the number of harmonics increases 
proportionally to $p^3$, and, the spectrum of emission
acquires the form of a synchrotron radiation rather than an 
undulator radiation.
In \cite{Korol98,Korol99} it was demonstrated that in a crystalline
undulator both possibilities can be realized.
However, to stay away from the synchrotron limit, $p\rightarrow \infty$,
it is desirable to consider moderate values of the undulator parameters 
$p\sim 1$.
This condition, accompanied with the inequality 
(\ref{Condition.2})
ensures that the spectrum of radiation formed in the undulator
will be presented by several powerful, narrow and well-separated peaks.
Taking into account that the undulator parameter is proportional to
the relativistic factor, 
which in ultra-relativistic case is much larger than one (typically,
$\gamma = 10^2\dots 10^4$ in crystalline undulators), 
the condition $p\sim 1$ results in a strong inequality 
$\lambda \gg a$ \cite{Korol98,Korol99}.
Combined with the large-amplitude regime, which is explained in section
\ref{LargeAndSmallAmpl}, we found that three quantities, $d$,
$a$ and $\lambda$, which characterize the crystalline undulator
must satisfy the following strong double inequality:
\begin{eqnarray}
d \ll a \ll \lambda.
\label{Condition.3}
\end{eqnarray}
This condition, clearly stated in \cite{Korol98,Korol99} and used
by us in further publications, together with the conditions
(\ref{Condition.1}) and (\ref{Condition.2}), are essential
for the effective operation of the crystalline undulator.
Let us note, that the last inequality in (\ref{Condition.3}) ensures, also,
the deformation of a crystal is the elastic one and does not
destroy the crystalline structure.

The periodic bending of the crystal can be achieved either dynamically or
statically. 
In \cite{Korol98}-\cite{Luederitz01} the main focus of our studies 
was made on the dynamic bending by means of a high-amplitude ($a\gg d$) 
transverse acoustic wave (AW), 
although the possibility of the static bending was pointed out as well.
The general formalism, developed in these papers for the description of 
the crystalline undulator and the radiation formed in it,  
does not depend on the method of the crystal bending. 
The analysis, which was performed in the cited papers, 
allowed us to establish the ranges of the AW amplitude and frequency
within which the conditions (\ref{Condition.1}), (\ref{Condition.2}) 
and (\ref{Condition.3}) as well as other very important conditions
(which are described below in this section) are fulfilled.
This was done for different types of projectiles,
wide ranges of their energies and for a variety of crystals. 
Thus, we proved theoretically the feasibility of the construction of
a micro-undulator by means of a periodically bent crystal and 
the possibility of generation of spontaneous and stimulated photon 
emission in such a system. 

The monochromatic transverse AW (either standing or
running) of the large amplitude, transmitted along a crystallographic plane,
allows to achieve a harmonic shape $y(z)=a\sin\Bigl(2\pi z/\lambda\Bigr)$
for the midplane.
One of the possibilities which can be used to do this 
is to place a piezo sample atop the crystal and to generate the radio 
frequency AW to excite the oscillations.
The important feature of the dynamic scheme of the crystal bending
is that the time period of the AW  must exceed greatly the time 
of flight $\tau = L/c$ of a bunch of particles through the crystal. 
Then, on the time scale of $\tau$, the shape of the crystal bending 
doesn't change, so that all particles of the bunch channel inside 
the same undulator. 
Thus, for the AW frequencies $\nu \le 100$ MHz, one gets $L \ll 70$ cm, 
which is more  than well-fulfilled for any realistic $L$-value
(for more details see \cite{Korol99}).

The idea to create an undulator by transmitting 
a high-amplitude transverse AW through a crystalline structure 
was mentioned for the first time in 
\cite{KaplinPlotnikovVorobev1980,BaryshevskyDubovskayaGrubich1980}
(see, also, \cite{Baryshevsky_Kniga}).
However, we state that in these papers neither the  feasibility of the 
crystalline undulator nor its theory were developed in a complete
and adequate form, because not all essential regimes, conditions and 
limitations were understood and elucidated. 
As a result, few statements and estimates, made in  
\cite{KaplinPlotnikovVorobev1980,BaryshevskyDubovskayaGrubich1980},
turn out to be not quite correct, as it is shown in sections 
\ref{DechannnelingAttenuation} and \ref{AllConditions}.
This is the reason why the idea of the crystalline undulator
based on the action of AW has not been attracting attention during
the period from 1980 till late 90's, when our first publications
\cite{Korol98,Korol99} appeared.

There were other papers
\cite{IkeziLinLiuOhkawa1984,BogaczKetterson1986,
Armyane1986,AmatuniElbakyan1988,MkrtchyanEtal1988,Dedkov94}, 
published prior to our first works \cite{Korol98}, 
devoted to the problem of radiation by 
ultra-relativistic particles (electrons and positrons) channeled in
the crystal in the presence of ultrasonic wave.
However, in all these publications the low-amplitude
regime $a < d$ (and, in some cases, the limit $a \ll d$) was
discussed.
Therefore, the main focus of these studies was made on various 
aspects of the influence of the additional undulator-type
oscillations of the particles on the spectrum of the channeling
radiation rather than on the properties of the crystalline undulator.

The important feature of a dynamically bent crystal by means of an
AW is that it allows  to consider an undulator with the parameters
$N$ and $p$  varying over a wide range, which is determined not
only by the projectile's energy but also by the AW frequency and
amplitude. 
The latter two quantities can easily be tuned resulting in the
possibility of varying significantly the intensity and shape of the
angular distribution of the radiation
(for a detailed discussion and an number of concrete examples 
see \cite{Korol98,Korol99}).
As demonstrated in \cite{Korol98,Korol99}, the parameters of 
the crystalline undulator based on the AW are inaccessible in conventional
undulators, where the periodicity of the motion of charged particles 
in achieved by applying periodic magnetic fields or the laser field 
\cite{ColsonFELs,Kniga,Alferov}.

The advantage of the static channel is that its parameters are fixed
and thus, the projectile moves along the fixed trajectory as well.
To calculate the characteristics of the emitted radiation one needs
to know only the number of the periods and the local curvature radius.
The disadvantage is that when fixing the number of periods 
the parameters of the system can be varied only by changing the energy
of the particle.  This makes the photon generation less tuneable.

For the first time the idea of a static crystalline undulator was proposed in
\cite{IkeziLinLiuOhkawa1984,BogaczKetterson1986} and implied the use of
a superlattice made of two constituents, which have different, but close,
lattice spacings. 
However, these papers were devoted only to the study of the low-amplitude
regime $a < d$.
As mentioned above, in this limit the intensive undulator radiation
does not appear.

It is feasible, by means of modern technology (like molecular beam 
epitaxy or chemical vapor deposition, see the references in 
\cite{Breese97,Sterzel2003}), to grow the crystal with its channels 
been statically bent according to a particular pattern. 
The usage of static methods to produce periodically bent 
crystals with $a\gg d$ was initially suggested in \cite{Korol98,Korol99} 
and later was discussed in \cite{MikkelsenUggerhoj2000}, where 
the idea to construct a crystalline undulator
based on graded composition strained layers was proposed.
Earlier, the same idea was exploited in \cite{SchaferGreiner1991}, where
the possibility to create a crystalline lens for the focusing of a beam of 
charged particles through a bent crystal was discussed.
Experimentally, the possibility of a 3 MeV proton beam bending by the 
Si$_{1-x}$Ge$_x$  graded composition strained layers was 
demonstrated in \cite{Breese97}.

In our papers \cite{Erevan01,Darmstadt01} we developed further the
ideas of \cite{MikkelsenUggerhoj2000,SchaferGreiner1991,Breese97} 
and demonstrated, for the first time, that it is possible to obtain 
periodically bent channels with {\it arbitrary} shapes $y(z)$.

In particular, it was described in detail, that to obtain a pure sine form of 
the channel profile, one starts with a pure silicon substrate and 
adds $\mathrm{Si}_{1-x}\mathrm{Ge}_x$ 
layers with continuously increasing Ge content. 
This results in bending of the (110) channels in the direction of the (100) 
channels. 
The periodicity of the shape requires the change of the direction of 
the bending toward the (010) channels. 
This, in turn, can be achieved by reducing $x$ until it reaches 0. 
The last (within the first period) crystal layer consists of pure
silicon, so that the second period can be built up on top of the first
in the same manner. 
To be captured by the bent channel, the positron
beam should be directed towards the (110) channel of the substrate.
The crystal strain is strongest after half a period, when the
germanium content reaches its maximum. 
The thickness of the layers
corresponding to half a period must be smaller than the critical
thickness $h_c$ \cite{Breese97}. 

In \cite{Erevan01,Darmstadt01} we developed the formalism and carried out 
the corresponding calculations which demonstrated, how one can 
built up a crystal, the channels of which are bent periodically
with arbitrary shapes $y(z)$.
 
In the papers by Avakian \etal 
\cite{AvakianAvetyanIspirianMelikyan2001_2,AvakianAvetyanIspirianMelikyan2002},
published within the same time interval as our papers 
\cite{Erevan01,Darmstadt01},
somewhat different approach, based on the use of graded composition strained
superlattices  $\mathrm{Si}_{1-x}\mathrm{Ge}_x$, 
for constructing the sine profile of crystal channels was described.
In the more recent publication \cite{AvakianAvetyanIspirianMelikyan2003} of 
this group the authors proceeded further with the study of 
the possibility to construct crystalline undulators  by means of
gradient crystals.
We want to point out that in this paper, which appeared after 
\cite{Erevan01,Darmstadt01}, the authors completely omitted the 
citation of our papers.
Indeed, the publication \cite{Darmstadt01} was ignored at all,
whereas the citation of \cite{Erevan01} 
(labeled as Ref. [7] in \cite{AvakianAvetyanIspirianMelikyan2003}) 
was done in an extremely negligent and misleading way which excludes any 
possibility for a reader to find our paper.  
Moreover, in the introductory part of their paper (p. 496) 
the authors of \cite{AvakianAvetyanIspirianMelikyan2003},
when reviewing `\dots a few proposals [2-5] for constructing micro-crystalline 
undulators\dots',
found it possible to include \cite{MikkelsenUggerhoj2000} and their
own publications 
\cite{AvakianGevorgianIspirianIspirian2001,
AvakianAvetyanIspirianMelikyan2001_2,
AvakianAvetyanIspirianMelikyan2002} in the citation list
 but to ignore our papers \cite{Korol98}-\cite{Darmstadt01}.
This very selective and misleading style of citation one finds in 
all publications of this group, see Refs. 
\cite{AvakianGevorgianIspirianIspirian2001,
AvakianAvetyanIspirianMelikyan2001_2,
AvakianAvetyanIspirianMelikyan2002,
AvakianAvetyanIspirianMelikyan2003}.
We find this style as totally unacceptable,
especially in the field which has become studied intensively 
{\it after} our first papers \cite{Korol98,Korol99}.
Unfortunately, this is not the only example of a selective
citation. 
It refers also to the papers by Bellucci \etal 
\cite{BellucciEtal2003,BellucciEtal2003_archive,Bellucci2003} 
and to a very recent paper 
\cite{KorhmazyanKorhmazyanBabadjanyan2004}, 
where the authors claimed that `the theory of radiation in micro-undulators
is developed' (see the abstract).

The periodic bending of crystallographic planes can also be achieved by
making regular defects either in the crystal volume or on its surface 
\cite{Romanov00}. 
Then, the crystalline planes in the vicinity of the defects become 
periodically bent. 
The practical realization of this idea was achieved in \cite{BellucciEtal2003} 
by giving periodic micro-scratches to one face of a silicon
crystal by means of a diamond blade.
Paying tribute to the fact that this was the first attempt of the 
actual construction of the micro-undulator, we, nevertheless,
want to point out that the introductory part of the paper 
\cite{BellucciEtal2003}
significantly misinterprets the history of the crystalline undulator idea.
Moreover, at the stage of preparation of the subsequent paper 
\cite{Bellucci2003} one of the authors of \cite{BellucciEtal2003} 
was aware of our work in this field, but found it possible to himself 
not to include our papers in the citation list.

\subsubsection{Restrictions due to the dechanneling effect 
and the photon attenuation.}
\label{DechannnelingAttenuation}

As was pointed out in \cite{Korol98,Korol99,Dechan01,Denton00,Durham00},
two physical phenomena, the dechanneling effect and the photon attenuation, 
lead to severe limitation of the allowed values 
of the crystalline undulator.

If the  dechanneling effect is neglected, one may unrestrictedly
increase the intensity of the undulator radiation by considering larger
$N$-values. 
In reality, random scattering of the channeling particle by the 
electrons and nuclei of the crystal leads to a gradual increase of 
the particle energy associated with the transverse oscillations 
in the channel. 
As a result, the transverse energy at some distance from the
entrance point exceeds the depth of the interplanar potential well,
and the particle leaves the channel. 
Consequently, the volume density $n(z)$ of the channeled particles 
decreases with the penetration distance, $z$, and, roughly, satisfies the
exponential decay law \cite{BiryukovChesnokovKotovBook}
\begin{equation}
n(z) = n(0)\, \exp(-z/ \Ld),
\label{DechannnelingAttenuation.2}
\end{equation}
where $n(0)$ is the volume density at the entrance.
The dechanneling length $\Ld(C)$, which is dependent on the parameter
$C$ (see (\ref{Condition.1})), depends also on the particle energy, 
mass and charge, on the parameters of the channel 
(its width and the distribution of electron charge in the channel), 
and on the charge of the crystal nuclei.

The dechanneling phenomenon introduces a natural upper limit on the 
length of a crystalline undulator: $L \leq \Ld(C)$.
Indeed, one can consider the limit $L \gg \Ld(C)$. 
However, as it was demonstrated in \cite{Dechan01,Denton00},
the intensity of the undulator radiation in this case is not 
defined by the expected number of the undulator periods $L/\lambda$ but
rather is  formed in the undulator of the effective length $\Ld(C)$.
Therefore, it is important to carry out realistic estimates of the quantity
$\Ld(C)$ in order to understand what are the limitations on the parameters
of the undulator imposed by dechanneling.  

In a model approach, presented in \cite{Korol99}, 
we used  the following expression 
for the dechanneling length in a periodically bent crystal:
\begin{eqnarray} 
\Ld(C) = (1-C)^2\, \Ld(0)
\label{DechannnelingAttenuation.3}
\end{eqnarray} 
where $\Ld(0)$ is the dechanneling length in a straight crystal.
For a positively charged projectile a good estimate for $\Ld(0)$ is 
\cite{Dechan01,BiryukovChesnokovKotovBook}:
\begin{eqnarray} 
\Ld(0) 
= 
\gamma\, {256 \over 9\pi^2}\,
{M\over Z}\, 
{a_{\rm TF} \over r_{0}}\,
{d \over \coul}
\label{DechannnelingAttenuation.4}
\end{eqnarray} 
where $r_{0}=2.8\times 10^{-13}$ cm is the electron classical radius,
$Z$, $M$ are the charge and the mass of a projectile 
measured in units of elementary charge and electron mass,
$a_{\rm TF}$ is the Thomas-Fermi atomic radius.
The quantity $\coul$  stands for a 'Coulomb logarithm' 
which characterizes the ionization losses of an
ultra-relativistic particle in amorphous media 
(see e.g. \cite{Land4,Sternheimer}):
\begin{eqnarray} 
\coul
=
\cases{
\ln{2\E\over I} -1
&for a heavy projectile\\
\ln{\sqrt{2}\E\over \gamma^{1/2}I} -23/24
&for $e^{+}$\\
}
\label{DechannnelingAttenuation.5}
\end{eqnarray} 
with $I$ being the mean ionization potential of an atom in the crystal.

In channeling theory it is more common 
\cite{BaurichterBiinoEtal2000,BiryukovChesnokovKotovBook}
to relate $\Ld(0)$ to the  parameter $pv$ ($p$ is the momentum of a projectile,
and in the ultra-relativistic case $pv\approx \E$) rather than to express it
via the relativistic factor $\gamma$ as in (\ref{DechannnelingAttenuation.4}).
However, explicit dependence of the dechanneling length on $\gamma$ 
is more convenient when discussing the parameters of the 
crystalline undulator and its radiation \cite{Korol99,Dechan01}.

\begin{figure}
\begin{center}
\epsfig{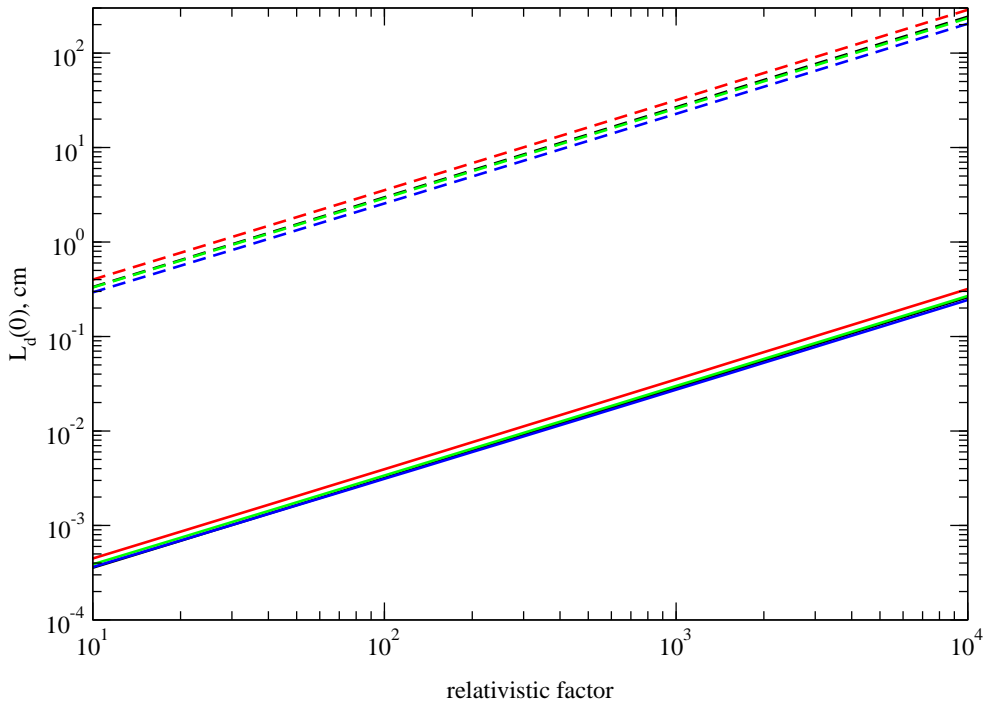}
\caption
{Dependence of $\Ld(0)$ on relativistic
factor $\gamma$ calculated from 
(\protect\ref{DechannnelingAttenuation.4})
-(\protect\ref{DechannnelingAttenuation.5}) for a positron (solid lines) 
and a proton (dashed lines) and for (110) channels in various crystals:
the black lines stand for $C$, 
the red lines - $Si$,
the green lines - $Ge$,
the blue lines - $W$.
}
\label{figure2_jpg}
\end{center}
\end{figure}

The dependences $\Ld(0)$ on $\gamma$ for a positron and a proton 
are illustrated by figure \ref{figure2_jpg}.
It is seen that in the case of a positron (the solid curves) the quantity
$\Ld(0)$ varies within $5\times10^{-4}\dots 0.3$ cm for $\gamma$
within $10\dots 10^4$.
The dechanneling length of a positron with energy within the GeV range
($\gamma\sim 10^4$) does not exceed several millimeters.
For a proton  (the dashed curves) of the same $\gamma$ the magnitude  
of $\Ld(0)$ is enhanced by the factor $\approx 10^3$.
This is largely due to the factor $M\approx 2000$ 
(see (\ref{DechannnelingAttenuation.4})).
Some additional correction originates from the difference of the Coulomb
logarithms, equation (\ref{DechannnelingAttenuation.5}).
To obtain the values of $\Ld(0)$ for a heavy ion one can multiply the
dashed curves by the factor $\approx A/Z\approx 2.5$. 

Using the results presented in figure \ref{figure2_jpg}
one can write the following estimate of the dechanneling 
length for various projectiles \cite{KorolSolovyovGreiner2003_2}:
\begin{eqnarray} 
\Ld(C)
\sim
(1-C)^2\,\gamma\times
\cases{
(2.5\dots 5)\times 10^{-5}\  \mbox{cm}   & \mbox{for}\ $e^{+}$\\
(0.05\dots0.1)\,  \mbox{cm}              & \mbox{for}\ $p$\\
(0.1\dots0.25)\,  \mbox{cm}              & \mbox{for a heavy ion}
}
\label{Condition.Ld}
\end{eqnarray} 

For heavy projectiles formulae 
(\ref{DechannnelingAttenuation.4})-(\ref{DechannnelingAttenuation.5}) 
are in good agreement with measured values of $\Ld(0)$ in a wide range
of $\gamma$ \cite{BaurichterBiinoEtal2000,BiryukovChesnokovKotovBook}.
In the case of a positron channeling in \cite{Dechan01} we tested 
equations (\ref{DechannnelingAttenuation.3}) 
and (\ref{DechannnelingAttenuation.4})  
against more rigorous calculation of the dechanneling lengths 
based on the simulation procedure of the positron channeling in 
straight and periodically bent crystals.
The approach developed and described in detail in  \cite{Dechan01} 
is based on the simulation 
of the trajectories and the  dechanneling process 
of an ultra-relativistic  positron.
This was done by solving the three-dimensional equations
of motion which account for: 
(i) the interplanar potential; 
(ii) the centrifugal potential due to the crystal bending;
(iii) the radiative damping force;
(iv) the stochastic force due to the random scattering of projectile
by lattice electrons and nuclei.
Note that the radiation damping force
becomes very significant at sufficiently large energies of positrons
(see section \ref{RadiativeLoss} for more details).
Simultaneously with simulating the trajectories of the channeled
particles, 
we calculated the total spectrum of the radiation, including
its undulator and channeling parts. 
This was done for all trajectories, including those which corresponded to
the over-barrier particles.
Such a study was carried out for the first time in \cite{TotSpect00,Dechan01}.
We point out that the referencing to these papers of ours
was not made in the later publications by Bellucci \etal 
\cite{BellucciEtal2003,BellucciEtal2003_archive,Bellucci2003}, where
the development of a simular approach was mentioned.

\begin{figure}
\begin{center}
\epsfig{file=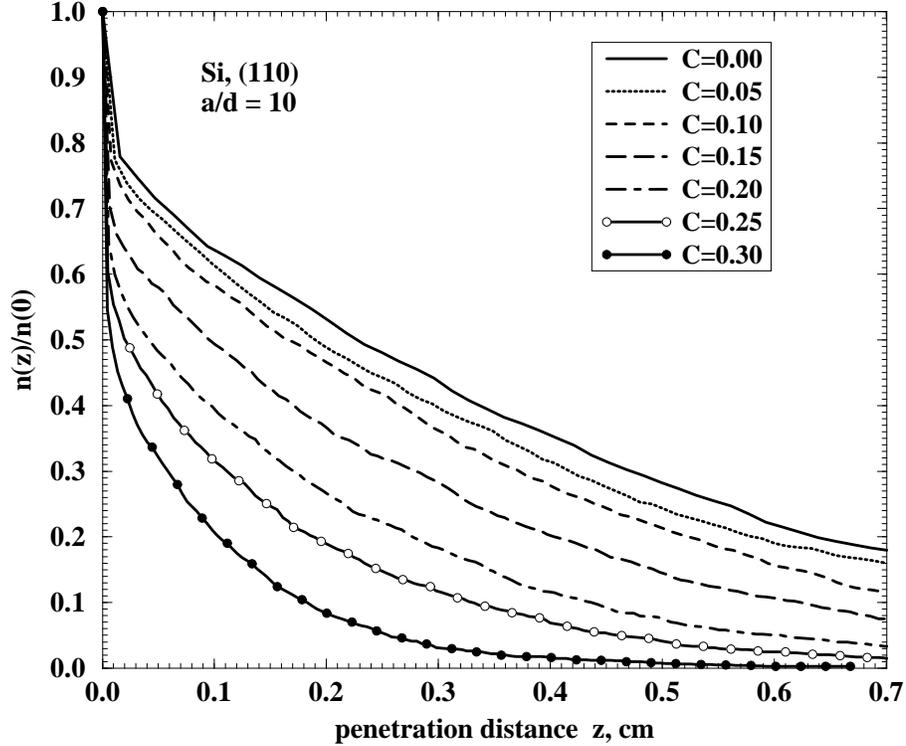,width=14cm, angle=0}
\caption{
The calculated dependences $n(z)/n(0)$ versus penetration 
distance $z$ for $5$ GeV positrons
channeling along the $(110)$ in Si crystal for various values
of the parameter $C$ \protect\cite{Dechan01}.
The $a/d$ ratio equals $10$.
The interplanar distance is $d=1.92$ \AA. 
}
\label{figure3_jpg}
\end{center}
\end{figure}

\begin{figure}
\begin{center}
\epsfig{file=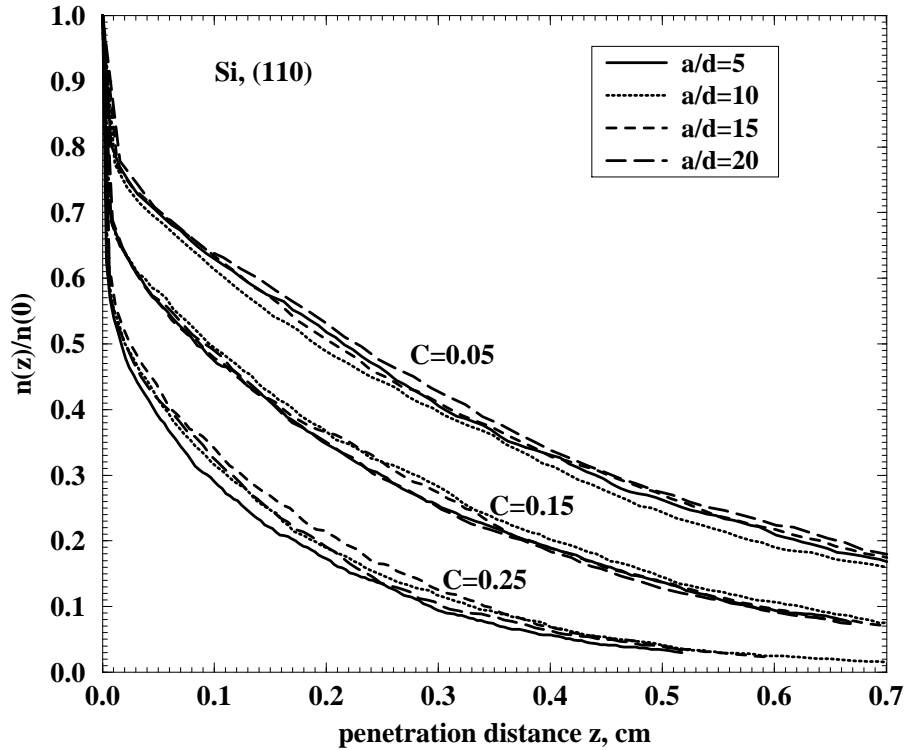,width=14cm, angle=0}
\caption{The calculated dependences $n(z)/n(0)$ versus penetration 
distance $z$ for $5$ GeV positrons
channeling along the $(110)$ plane in a Si crystal for various values
of the parameter $C$ and of the ratio $a/d$ \protect\cite{Dechan01}.  
}
\label{figure4_jpg}
\end{center}
\end{figure}

We analyzed the dependence of $\Ld$ on the energy of the projectile,
the type of crystal and crystallographic plane, the parameter
$C$ and the ratio $a/d$  (see  (\ref{Condition.1}) and
(\ref{Cond4})).
Some results of these calculations, corresponding to $5$ GeV positrons
channeling along the $(110)$ plane in a Si crystal, 
are presented in figures \ref{figure3_jpg} and  
\ref{figure4_jpg}.
\begin{table}
\caption{Dechanneling lengths for 5 GeV positron channeling
along the $(110)$ planes for various crystals and for various values
of the parameter $C$.   
The $a/d$ ratio equals $10$ except for the case $C=0$ 
(the straight channel).
The quantity $\Ld^c$ presents the results of our calculations
published in \protect\cite{Dechan01}.
$N_d^c=\Ld^c/\lambda$ is the corresponding 
number of the undulator periods,  
$\Ld^e$ is the dechanneling length estimated according to 
(\protect\ref{DechannnelingAttenuation.3})- 
(\protect\ref{DechannnelingAttenuation.5}), $N_d^e=\Ld^e/\lambda$.
For each $C$  the value of $\lambda$ was derived from
(\protect\ref{Condition.1}).
Other parameters are:
$\hbar\om_1$ is the energy of the first harmonic of the undulator
radiation for the forward
emission, $p$ is the undulator parameter.}
\begin{indented}
\item[]\begin{tabular}{@{}rrrrrrrrr}
\br
  $C$   &$\lambda$  &$\Ld^e$ & $\Ld^c$ &$N_d^e$&$N_d^c$&$\omega_1$&$p$\\
        & $\mu$m    &  cm    &  cm     &       &    &  MeV   & \\
\br
 0.00   &   -       &  0.312 &  0.463  &   -   &  - &   -  &-\\
 0.05   &  100.9    &   0.281 &  0.430  &   25  &39  &1.38& 1.08 \\
 0.10   &  77.1     &   0.253 &  0.393  &   32  &51  &1.42& 1.53 \\
 0.15   &  63.0     &   0.225 &  0.321  &   35  &51  &1.37& 1.87 \\
 0.20   &  54.5     &   0.200 &  0.223  &   36  &41  &1.31& 2.16 \\
 0.25   &  48.8     &   0.175 &  0.170  &   35  &35  &1.24& 2.42 \\
 0.30   &  44.5     &   0.153 &  0.102  &   34  &23  &1.18& 2.65 \\
\br
\end{tabular}
\end{indented}
\label{Table1}
\end{table}

In table \ref{Table1} the results of our calculation 
\protect\cite{Dechan01} of the dechanneling lengths are compared
with those estimated from 
(\protect\ref{DechannnelingAttenuation.3}).
It can be concluded that the approximate formula
(\ref{DechannnelingAttenuation.3}) adequately reproduces the values of 
$\Ld(C)$ for a positron in the range $C=0\dots 0.2$.
As noted in \cite{Dechan01}, some discrepancy between the calculated, 
$\Ld^c$, and the estimated $\Ld^e$, dechanneling lengths 
can be attributed to the fact that the quantity 
$\Ld(0)$, defined in (\ref{DechannnelingAttenuation.4}),
was obtained by using  the Lindhard planar
potential (see, e.g., \cite{Gemmell}) and, additionally, several simplifying
assumptions were made concerning the electron charge distribution in the 
channel (see  \cite{BiryukovChesnokovKotovBook}).
On the other hand, in \cite{Dechan01} the numerical procedure was based 
on the Moli\`ere approximation for both the potential and the charge 
distribution.
More discussion on the comparison of the model
(\ref{DechannnelingAttenuation.3}) 
and the data obtained numerically one finds in \cite{Dechan01}.
In the cited paper similar calculations were performed also 
for $Ge$ and $W$ crystals.

\begin{figure}
\begin{center}
\epsfig{file=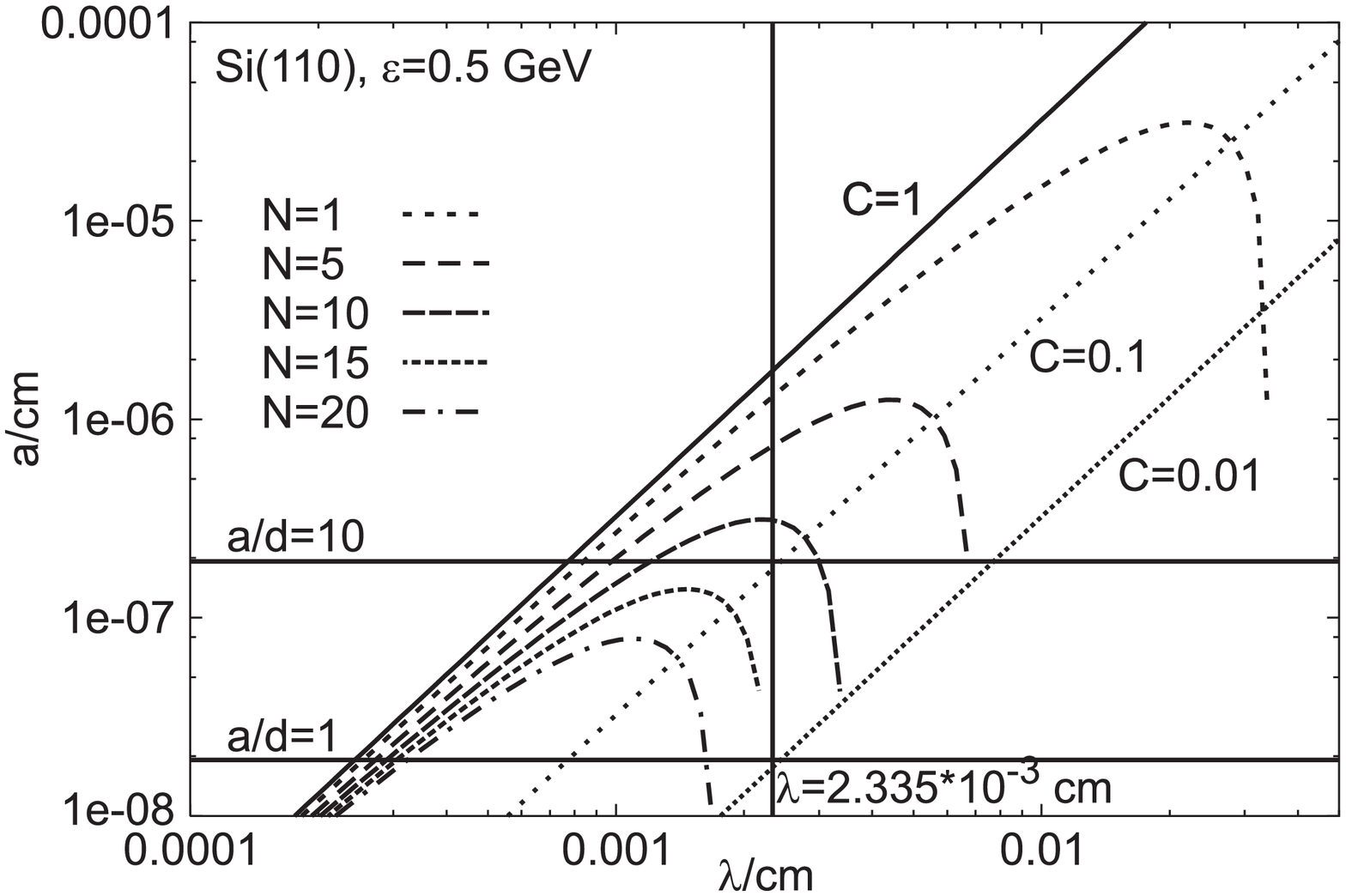,width=14cm, angle=0}
\caption{The range of parameters $a$ and $\lambda$ for a bent
Si(110) crystal at $\varepsilon=500$ MeV.
}
\label{figure5_jpg}
\end{center}
\end{figure}

Combining conditions (\ref{Condition.1}), (\ref{Condition.2}), 
 (\ref{Condition.3}) and
$L \leq \Ld(C)$
one can find, fixing a crystal and energy $\E$,
the allowed ranges of the parameters $a$ and $\lambda$  
\cite{Korol99,TotSpect00,Dechan01}.
Figure \ref{figure5_jpg} illustrates this for 
the case of $\E=0.5$ GeV positrons planar channeling 
in Si along the (110) crystallographic planes.  
The diagonal straight lines correspond to various values of the 
parameter $C$.  
The curved lines correspond to various values of the number of
undulator periods $N$ related to the dechanneling length $\Ld$
through $N=\Ld/\lambda$.  
The horizontal lines mark the values of the
amplitude equal to $d$ and to $10\,d$, where 
$d=1.92$ \AA \ is the interplanar spacing for  the (110) planes in Si.
The vertical line marks the value $\lambda = 2.335\times 10^{-3}$ cm, 
for which the spectra (see section \ref{CrystUndRad}) were calculated.

Figure \ref{figure5_jpg} illustrates that the restrictions, 
imposed by the dechanneling effect on the length of a crystalline undulator,
are very severe, especially in the case of a projectile positron.
Therefore, the influence of the dechanneling must be studied very 
accurately in order to produce realistic predictions on the
parameters of the crystalline undulator and the undulator emission.
For the first time this comprehensive analysis was performed in
\cite{Korol98,Korol99,TotSpect00,Dechan01}.
In the earlier papers the dechanneling effect was either completely
ignored (see 
\cite{BaryshevskyDubovskayaGrubich1980,IkeziLinLiuOhkawa1984,
BogaczKetterson1986,Dedkov94})
or its role was estimated erroneously, as in 
\cite{KaplinPlotnikovVorobev1980}.
In particular, in the latter paper, where the authors did 
not explicitely distinguish the electron and positron channeling,
the following comment was made:
`We recall that the channeling length in centimeters is approximately
$L_0=\E$ (GeV)\dots' 
(see page 650 in \cite{KaplinPlotnikovVorobev1980}).
This is absolutely incorrect.
Indeed, as it is seen from 
figures \ref{figure2_jpg}-\ref{figure5_jpg} and table \ref{Table1},
the dechanneling lengths for a positron are, at least, an order
of magnitude less, not mentioning the electron case, when
the values of $\Ld$ are even more lower. 
As a result, the estimated values of $\lambda$, $a$ and $\om$,
which were suggested in
\cite{KaplinPlotnikovVorobev1980,BaryshevskyDubovskayaGrubich1980},
for a crystalline undulator based on a $\varepsilon = 1$ GeV
positron channeling in Si, 
lay far away from the regions
allowed for feasible crystalline undulators.
In section \ref{AllConditions} below we discuss in more detail
the parameters suggested in
\cite{KaplinPlotnikovVorobev1980,BaryshevskyDubovskayaGrubich1980}.

The propagation of photons emitted in a crystalline undulator
is strongly influenced by a variety of processes occuring in a crystal.
These are the atomic and the nuclear photoeffects, the coherent and 
incoherent scattering on electrons and nuclei, the
electron-positron pair production (in the case of high energy
photons).
All these processes lead to the decrease in the intensity of the photon
flux as it propagates through the crystal:
\begin{eqnarray}
I(z) = I(0)\, \exp(-z/ L_a(\om))
\label{DechannnelingAttenuation.6}
\end{eqnarray}
where $I(0)$ is the initial intensity, 
$I(z)$ is that which remains after traversal of the distance $z$.
A quantitative parameter, which we introduced in 
(\ref{DechannnelingAttenuation.6}) to account for all these effects, 
can be called the attenuation length,  $L_a(\om)$. 
It is related to the mass attenuation coefficient $\mu(\omega)$ as
$L_a(\om)=1/\mu(\omega)$ \cite{Henke,Hubbel,ParticleDataGroup2002}.
Thus, $L_a(\om)$ defines the scale within which the
intensity of a photon flux decreases by a factor of $e$. 

\begin{figure}
\begin{center}
\epsfig{file=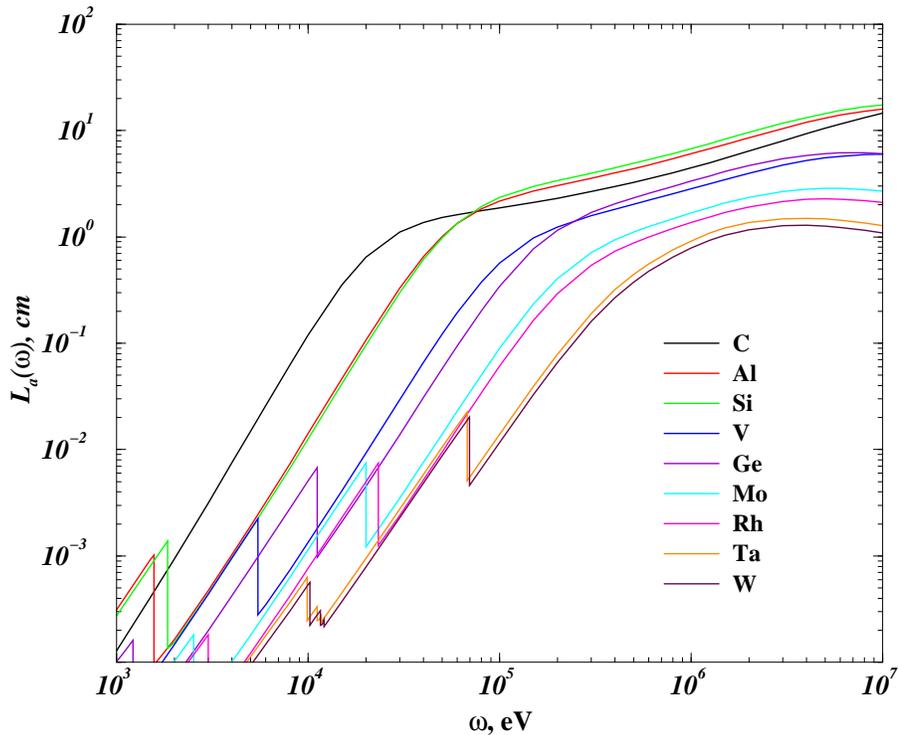 ,width=12cm}
\vspace*{-0.8cm}
\caption{Attenuation length $L_a(\om)=1/\mu(\omega)$ versus 
the photon energy for various crystals, as indicated.
The data on $\mu(\omega)$  are taken from \protect\cite{Hubbel}.}
\label{figure6_jpg}
\end{center}
\end{figure}

The mass attenuation coefficients are tabulated for all elements
and for a wide range of photon frequencies.
Figure \ref{figure6_jpg} represents the dependences $L_a(\om)$ for a 
variety of crystals over the broad range of photon energies.
For photon energies $30\ \mbox{eV} < \hbar\om < 1$ keV the mass 
attenuation coefficients 
(which are mainly due to the atomic photoeffect) can be found in
\cite{Henke}.
The corresponding values of $L_a(\om)$ are lower than the minimum values 
in figure \ref{figure6_jpg}.
However, for $\hbar\om \ll I_0$ (where $I_0$ is the ionization potential of 
the crystal atom, and $I_0 \leq 10$ eV for most of crystals)
there is no photoabsorption and, therefore, the attenuation is 
defined solely by elastic photon scattering, i.e. is comparatively weak.
Using these arguments and the data presented in figure \ref{figure6_jpg}
can be summarized in the form convenient  for a quick estimation 
of $L_a(\om)$:
\begin{eqnarray}
L_a(\om) 
\approx
\cases{
\infty                  & for $\hbar\om \ll I_o \leq 10$  eV\\
\ll 10^{-2}\ \mbox{cm}  & for $\hbar\om=10^{-2}\dots 10$  keV\\
=0.01\dots 10\ \mbox{cm}& for $\hbar\om > 10$ keV
}
\label{Condition.La}
\end{eqnarray}

The quantities $\Ld(C)$ and $L_a(\om)$ introduced above in this
section define the effective upper limit of the crystal length $L$
which can be used to calculate the number
of undulator periods $N$ \cite{Korol99}:
\begin{eqnarray}
L < \min\left[\Ld(C),L_a(\om)\right]
\label{Conditions.5}
\end{eqnarray}


\subsection{Energy losses and shape of crystalline undulator.}
\label{RadiativeLoss}

\begin{figure}
\begin{center}
\epsfig{file=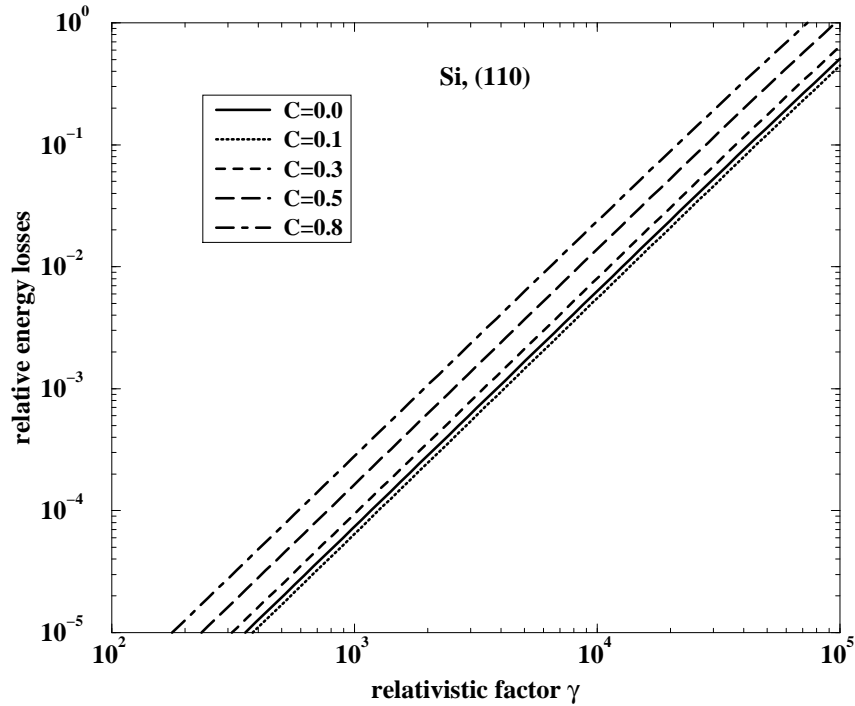 ,width=13cm}
\caption{Relative energy losses for the crystal length $L=\Ld(C)$ 
versus relativistic factor $\gamma$. The curves correspond to different
values of the parameter $C$ (see  (\protect\ref{Condition.1})) 
\protect\cite{EnLoss00}.}
\label{figure7_jpg}
\end{center}
\end{figure}

The coherence of the radiation, emitted from similar parts of the 
trajectory of a particle in the crystalline undulator, takes
place if the energy of the channeling particle does not 
change noticeably during with the penetration distance, at least, on the scale
of the dechanneling length.
For ultra-relativistic projectiles the main source of energy losses
are the radiative losses \cite{Baier,Land4}.
Therefore, it is important to establish the range of energies of 
channeling particles for which the parameters of undulator radiation formed
in a perfect periodic crystalline structure are stable.
For the first time, the importance of the restrictive role of the 
radiative losses was realized in \cite{Korol99}.
Later, in \cite{EnLoss00}, we carried out a comprehensive theoretical
and numerical analysis of the radiative loss of energy, $\Delta \E$, 
of ultra-relativistic positrons channeling in crystalline undulators.
General formalism described in \cite{EnLoss00} is applicable for the
calculation of the total losses, which account for the contributions 
of both the undulator and the channeling radiation. 
We analyzed the relative importance of the two mechanisms
for various values of $a$, $\lambda$, and $\E$.
We established the ranges of energies for positrons, in which relative
radiative losses, $\Delta \E/\E$, are small (lower than 1 per cent) 
for a variety of crystals and crystallographic planes.
The results of these calculations are illustrated in figure \ref{figure7_jpg},
where the dependences $\Delta \E/\E$ versus relativistic factor $\gamma$
are presented for a positron channeling in (110) channel of Si
and for several values of the parameter $C$.
The analysis, performed in \cite{EnLoss00} for the crystals
LiH, C, Si, Ge, Fe and  W, demonstrated that 
for a perfectly shaped crystalline undulator (i.e., the one
in which the midplane is modulated as $y(z)=a\sin(2\pi z/\lambda)$,
see figure \ref{figure7_jpg})
the radiative energy losses become large if the
initial energy of the positron bunch is $\E > 10$ Gev. 
For lower energies of positrons, when the relativistic
factor satisfies the the inequality 
\begin{eqnarray}
\gamma \leq 10^4,
\label{Conditions.4}
\end{eqnarray}
the radiative losses are small, $\Delta \E < 0.01 \E$.

The condition (\ref{Conditions.4}) establishes the upper limit 
of positron energies which is meaningful to use to generate the 
stable undulator radiation in the ideal crystalline undulator.
In the high-energy regime, when $\E>10$ Gev, the gradual decrease of the 
positron energy strongly influences the stability of the parameters
of the undulator radiation.
However, in \cite{Erevan01,Darmstadt01} we demonstrated,  for the first time,
that the coherence and the monochromaticity of the undulator radiation 
in the high-energy regime can be maintained if the amplitude 
and the period of the bent channel are 
made dependent on the penetration distance $z$, i.e. $a=a(z)$, and
$\lambda=\lambda(z)$.
We derived the equations for these dependences and found the corresponding
solutions.
The method of preparation of the crystals the midplanes of which are
shaped as $y(z)=a(z)\sin(2\pi z/\lambda(z))$ was described in detail.
This method is based on the crystal growing by means of molecular beam
epitaxy or chemical vapor deposition of a crystal with graded strained 
layers \cite{Breese97}. 
As an example, we considered a pure silicon substrate on which a
Si$_{1-x}$Ge$_x$ layers are added. 
Here $x=x(z)$ is the germanium content in the layer, and it is 
varied during the  growing process in order to achieve the desired 
shape of the channels.

\subsection{Feasibility of a crystalline undulator} 
\label{AllConditions}

Let us summarize all the conditions 
which must be fulfilled in order to treat a crystalline undulator
as a feasible scheme for devising on its basis new sources of
electromagnetic radiation.
These conditions are:
\begin{eqnarray}
\fl
\cases{
C =(2\pi)^2{\E \over q U_{\max}^{\prime}} {a\over \lambda^2} \ll 1
& stable channeling
\\
d \ll a \ll \lambda
&
large-amplitude regime
\\
N = {L \over \lambda} \gg 1
&
large number of undulator periods
\\
L < \min\left[\Ld(C),L_a(\om)\right]
&
account for the dechanneling and photon attenuation
\\
{\Delta \E \over \E} \ll 1
&
low radiative losses 
}
\label{AllConditions.1}
\end{eqnarray}
As a supplement to this system one must account for the formulae 
(\ref{Om0andP}) and (\ref{harmonics.1}), which define the parameters 
of the undulator and the frequencies of the undulator radiation.

Provided all conditions (\ref{AllConditions.1}) are met 
for a positively charged particle channeling through a periodically bent 
crystal then
\begin{itemize}
\item
within the length $L$ the particle experiences stable planar channeling 
between two adjacent crystallographic planes,
\item
the characteristic frequencies of the undulator radiation and the ordinary 
channeling  radiation are well separated,
\item
the intensity of the undulator radiation 
is essentially higher than that of the ordinary 
channeling  radiation,
\item
the emission spectrum is stable towards the radiative losses of the
particle.
\end{itemize}

For each type of the projectile and its energy, for 
a given crystal and crystallographic plane 
the analysis of the system (\ref{AllConditions.1})
is to be carried out in order to establish the ranges 
of $a$, $\lambda$ and $\om$ within which the operation of the
crystalline undulator is possible.

Most of these important conditions were realized
and carefully investigated for the first time in 
\cite{Korol98}-\cite{Darmstadt01},
where the realistic numerical calculations of the characteristics
of the radiation formed in crystalline undulator were performed as well.
We consider the set of analytical and numerical results obtained by us
in the cited papers as a proof of the statement 
that the scheme illustrated in figure  \ref{figure1_jpg} can be 
transformed from the stage of a  purely academic idea 
up to an observable effect and an operating device.

For a positron channeling, in particular,  we found the 
the optimal regime in which the spontaneous undulator radiation is most 
stable and intensive, and demonstrated that this regime is realistic.
This regime is characterized by the following ranges of the parameters:
$\gamma=(1\dots 10)\times 10^3$, $a/d=10\dots50$,
$C=0.01\dots0.2$, 
which are common for all the crystals which we have investigated.
These ranges ensure that the energy of the first harmonic $\om_1$
(see (\ref{harmonics.1}))
lies within the interval $50 \dots 150$ keV and the length of the 
undulator can be taken equal to the dechanneling length because of the
inequality $\Ld(C) < L_a(\om)$.

The importance of exactly this regime of operation of the crystalline
undulator was later realized by other authors.
In particular, in recent publications by Bellucci \etal
\cite{BellucciEtal2003,BellucciEtal2003_archive,Bellucci2003},
where the first practical realization of the crystalline undulator was
reported, the parameters chosen for a Si crystal were as follows:
$\E=0.5\dots 0.8$ GeV for a positron (i.e. $\gamma=(1\dots1.6)\times10^3$),
$a=20\dots 150$\, \AA (i.e. $a/d=10\dots 80$), $L=\Ld$.
These are exactly the values for which we predicted the strong
undulator effect.
However, in these papers, where 
the authors mention all the conditions (\ref{AllConditions.1}) and
stress their importance,
there is no proper reference to our works.
Instead, our paper \cite{Korol99}, labeled as Ref. [10] in 
\cite{BellucciEtal2003}, was cited as follows:
`With a strong world-wide attention to novel sources of 
radiation, there has been broad theoretical interest [4-12]
in compact crystalline undulators\dots' (page 034801-1 in the cited paper).
This was the only referencing to the paper \cite{Korol99}, in which we
clearly formulated, for the first time, 
the conditions (\ref{AllConditions.1})
and carried out a detailed analysis aimed to prove why this regime
is most realistic. 
None of it was done in the papers 
\cite{KaplinPlotnikovVorobev1980,BaryshevskyDubovskayaGrubich1980,
IkeziLinLiuOhkawa1984,BogaczKetterson1986,Dedkov94}
(labeled in \cite{BellucciEtal2003} as Refs. [4],[6],[7],[8] and [9],
correspondingly).
Moreover, we state that one will fail to construct a crystalline undulator
basing on the estimates presented in  
\cite{KaplinPlotnikovVorobev1980,BaryshevskyDubovskayaGrubich1980,
IkeziLinLiuOhkawa1984,BogaczKetterson1986,Dedkov94}.
In what follows  we carry out critical analysis
of the statements and the estimates made in the cited papers.

Historically, the paper by Kaplin \etal \cite{KaplinPlotnikovVorobev1980}
was the first one, where the idea of a crystalline undulator based
on the action of the transverse acoustic wave was presented.
However, a number of ambiguous or erroneous statements 
makes it impossible to accept the thesis that the concept of a crystalline
undulator was correctly described in this two-page paper.
To be precise in our critics, below we use the exact citations
taken from the English edition of \cite{KaplinPlotnikovVorobev1980}.
In the citations the italicizing is made by us.

Our first remark concerns the type of a projectile which the 
authors propose to use in the undulator.
The first paragraph of the paper contains:

\noindent
`Radiation by relativistic {\it electrons and positrons}, which
 occurs during channeling in single crystals, has been observed
experimentally and is being extensively studied at the present
time$^{1-4}$.'

\noindent
This is the {\it only} place in the text where the term 'positron' is used.
In the rest of the paper the projectile is called either 'a particle', or
a 'relativistic electron' as in one before the last paragraph of the paper
(page 651).
Thus, it is absolutely unclear to the reader, which particle is to be
used. For a positron it is possible to construct an undulator, however
if an electron is considered, then the rest of the paper does not make
any sense.

The concept of a periodically bent crystal and its parameters is formulated
as follows (page 650, right column):

\noindent
`Still higher intensity can be achieved by using instead of a uniformly 
curved crystal one deformed in such a way that the radiation from 
different portions of the particle trajectory adds coherently.
This can be accomplished by giving a crystalline plate a wavelike shape 
in such a way that the sagitta $A$ satisfies the relation
$4A\gamma/ \lambda_0 <1$ in relation to the quarter period $\lambda_0$
of the bending.
{\it For large values of the dechanneling depth $L_0$ this will provide a high radiated power from the
crystalline undulator (wiggler)}.
For rather thin crystalline plates with a simple bend one can produce
$\lambda_0\sim 4$ mm \dots 
{\it We recall that the channeling depth in
centimeters is approximately $L_0 = E$ (GeV)}, as follows from experiments.'

\noindent
Note, that no citation is made when referring to the experiments
which result in `$L_0$ (cm) = $E$ (GeV)'. 
For a positron (see section \ref{DechannnelingAttenuation}) this 
relation overestimates the dechanneling length by more than an order of
magnitude, for an electron it is even farther from the reality.
Therefore, the idea to construct an undulator for a positron with the 
period $\lambda = 4\lambda_0 = 1.6$ cm is absolutely unrealistic.

The parameters of the undulator based on the action of the acoustic wave
are presented in the left column on page 651:

\noindent
`To obtain radiation in the optical region in a transparent
crystal or to generate very hard $\gamma$ rays, it has been proposed
to use ultrasonic vibrations to deform the crystal lattice\dots
For example, one can obtain $\gamma$ rays with the energy up to
$\om =0.14-14$ MeV for $\E=1$ GeV and $\lambda_0=10-0.1$ $\mu$m.'

\noindent
Note, that none of the following characteristics, - 
the type of the projectile, the crystal, the acoustic wave amplitude
(in our notations 'sagitta A' is called 'amplitude $a$'),  are specified.
{\it Assuming} that the positron channeling is implied, let us
analyze the above mentioned values from the viewpoint of the condition 
for a stable channeling, equation (\ref{Condition.1})
(see also (\ref{AllConditions.1})).
The parameter $C$ can be written in the form:
$C \approx 40/\lambda^2\, (\E\, d/ q U_{\max}^{\prime})\,(a/d)$
with $\lambda$ in $\mu$m, $\E$ in GeV, $d$ in \, \AA, and 
$q \dUmax$ in GeV/cm.
Let us estimate the ratio $a/d$ for the range 
$\lambda = 4\lambda_0=0.4-40$ $\mu$m
and for (110) planes in  Si and W, for which
$d_{\rm Si}=1.92$\, \AA, $d_{\rm W}=2.24$\, \AA,  
$\left(q \dUmax\right)_{\rm Si}=6.9$ Gev/cm,
$\left(q \dUmax\right)_{\rm W}=57$ Gev/cm
\cite{Baier}.
For $\E=1$ GeV and the lowest $\lambda$-value one gets
$C\approx 250 (\E\, d/ q \dUmax)\,(a/d)$, which
means that, for both crystals, to satisfy the condition $C\ll 1$ it 
is necessary to consider $a \ll d$.
Thus, this is a low-amplitude regime, for which 
the intensity of the undulator radiation is negligibly small.
The upper limit of $\lambda$ is more realistic to ensure the
condition $C\ll 1$ for the amplitudes $a \gg d$.
However, this analysis is not performed by the authors.

Our final remark concerns the statement (the last
paragraph in the left column on page 651):

\noindent
`A lattice can be deformed elastically up to $A=1000$\, \AA \dots'

\noindent
This is true, but when referring to the crystalline undulator with the
amplitude $a=10^{-5}$ cm one has to supply the reader (and a potential
experimentalist) with the estimates of the corresponding values
of $\lambda$ and $N$. 
Let us carry out these estimates (note, this was {\it not} done in the 
paper).
The channeling condition (\ref{Condition.1})
can be written as follows:
\begin{eqnarray} 
\lambda =
{\lambda_{min} \over \sqrt{C}} > \lambda_{min},
\label{kaplin.1}
\end{eqnarray} 
where $\lambda_{\min}$ is the absolute minimum of $\lambda$ 
(for given $a$, $\E$ and a crystal) which corresponds to
$C=1$ (i.e. to the case when the dechanneling length 
$\Ld(C)$ effectively equals to zero, 
see (\ref{DechannnelingAttenuation.3})).
It is equal to
\begin{eqnarray} 
\lambda_{\min} =
2\pi\, \sqrt{a}
\left( {\E \over q\dUmax }\right)^{1/2}.
\label{kaplin.2}
\end{eqnarray} 
For a 1 GeV positron channeling in Si and W crystals along the (110) plane,
which is plane bent periodically with $a=10^{-5}$ cm,
the values of $\lambda_{\min}$ are: $7.6\times10^{-3}$ cm for Si
and  $7.5\times10^{-3}$ cm for W.
These values already exceed the upper limit of $40$ $\mu$m mentioned
by Kaplin \etal.
Choosing the length of the crystal to be equal to the dechanneling length
and using equation (\ref{DechannnelingAttenuation.3}) to estimate
$\Ld(C)$ one estimates the number of undulator periods
$N=\Ld(C)/\lambda=C^{1/2}(1-C)^2 \Ld(0)/\lambda_{\min}$.
The largest value of $N$ is achieved when $C=0.2$, giving
$C^{1/2}(1-C)^2 \approx 0.29$.
Hence, $N\leq N_{\max}=0.29\Ld(0)/\lambda_{\min}$.
Using formulae (\ref{DechannnelingAttenuation.4}) and 
(\ref{DechannnelingAttenuation.5}) one calculates the dechanneling
lengths in straight crystals: 
$\Ld(0) = 6.8\times10^{-2}$ cm $\Ld(0) = 3.9\times10^{-3}$ cm for W.
Finally, one derives that the `undulator' suggested in the
cited paper contains $N\leq 2.6$ periods in the case of Si,
and $N\leq 1.5$ for a tungsten crystal. 

Thus, because of the inconsistent and ambiguous character of the 
paper \cite{KaplinPlotnikovVorobev1980} we cannot agree with the statement,
the feasibility of a crystalline undulator was demonstrated in
this paper in a manner, sufficient to stimulate the experimental
study of the phenomenon.
 
None of the essential conditions, summarized in (\ref{AllConditions.1}),
were analyzed in  \cite{KaplinPlotnikovVorobev1980}.
For the first time such an analysis was carried out in 
\cite{Korol98,Korol99} and developed further in our 
subsequent publications.
In this connection we express disagreement with utterly
negligent and unbalanced style of citation adopted by Avakian \etal in 
\cite{AvakianGevorgianIspirianIspirian2001}
and other publications 
\cite{AvakianAvetyanIspirianMelikyan2001_2,
AvakianAvetyanIspirianMelikyan2002,
AvakianAvetyanIspirianMelikyan2003} by this group,
and by Bellucci \etal
\cite{BellucciEtal2003,BellucciEtal2003_archive,Bellucci2003}.

Much of our critics expressed above in connection with 
\cite{KaplinPlotnikovVorobev1980} refers also to the paper
by Baryshevsky \etal \cite{BaryshevskyDubovskayaGrubich1980}.
The main point of ours is: the concept of the crystalline undulator
based on the action of an acoustic wave was not convincely presented.
>From the text of the paper it is not at all clear 
what channeling regime, axial or planar, should be used.
The only reference to the regime is made in last part of the paper, 
on page 63, which is devoted to the quantum description of 
the spectral distribution of the undulator radiation. 
This part starts with the sentence:
 'Let us consider, for example, planar channeling'. 
The question on whether the axial 
channeling is also suitable for a crystalline undulator is left 
unanswered by the authors.
Neither is it clearly stated what type of a projectile
is considered.
Indeed, in all parts of the paper, where the formalism is presented,
the projectile is called as a 'particle'.
The reference to a positron is made in the introductory paragraph, where
the effect of channeling radiation is mentioned, and on page 62, where
the numerical estimates of the intensity of the undulator radiation 
are presented.
The limitations due to the dechanneling effect are not discussed.
As a consequence, the regime, for which the estimates are made,
hardly can be called the undulator one.
Indeed, on page 62 the ratio of the undulator to the channeling radiation
intensities is estimated for a 1 GeV positron channeled in Si 
({\it presumably}, the planar channeling is implied).
The amplitude of the acoustic wave (labeled as $r_{0\perp}^s$) is
chosen to be equal to $10^{-5}$ cm.
The period $\lambda$ is not explicitely written by the authors.
However, they indicate the frequency of the acoustic wave,
$f=10^7$ s$^{-1}$.
Hence, the reader can deduce that 
$\lambda = v/f=4.65\times10^{-2}$ cm, if taking the value
$v=4.65\times10^{5}$ cm/s for the sound velocity in Si 
\cite{Mason}.
The values of $\E$, $a$ and $\lambda$, together with 
the maximal gradient of the interplanar field
$\left(q \dUmax\right)_{\rm Si}=6.9$ Gev/cm \cite{Baier}, allows
one to calculate $C=2.65\times 10^{-2}$ (see (\ref{AllConditions.1})), and,
consequently, to estimate the dechanneling length 
$L_d(C) = 6.47\times 10^{-2}$ cm.
As a result, we find that the number of the undulator
periods in the suggested system is $N= 1.4$, which 
is not at all $N\gg 1$ as it is implied by the authors 
(this is explicitly accented by them in the remark in the line 
just below their equation (2) on page 62).
Another point of critics is that the classical formalism, used
to derive the equation (2), is applicable only for the dipole case, i.e.
when the undulator parameter is small, $p^2\ll 1$.
However, the estimates which are made refer to a strongly non-dipolar
regime: $p^2 = (2\pi\gamma a/\lambda)^2 = 7.3$.
As a consequence,  the estimate of the energy of the largest emitted
harmonic, carried out by the authors on page 62, is totally wrong.
Exactly in their regime the harmonics with low number will never
emerge from the crystal due to the photon attenuation.

Papers \cite{IkeziLinLiuOhkawa1984,BogaczKetterson1986,Dedkov94}.
considered {\it only}
the case of small amplitudes,  $a \ll d$, when discussing 
the channeling phenomenon in periodically bent crystalline
structures. 
As a result, in \cite{IkeziLinLiuOhkawa1984,Dedkov94} the attention was
paid not to the undulator radiation (the intensity of which is negligibly
small in the low-amplitude regime, see section \ref{CrystUndRad}), 
but to the influence of the periodicity of the channel bending on the
spectrum of the channeling radiation.
Similar studies were carried out in 
\cite{Armyane1986,AmatuniElbakyan1988,MkrtchyanEtal1988,GrigoryanEtal2001,
GrigoryanEtal2003}.
These effects are irrelevant from the viewpoint of the 
crystalline undulator problem discussed here.
Another issue, which we want to point out, is that the authors 
of \cite{IkeziLinLiuOhkawa1984,Dedkov94} did not distinguish between 
the cases of an electron and a positron channeling.
The limitations due to the dechanneling effect were not discussed.
In \cite{BogaczKetterson1986} the idea of using a superlattice 
(or a crystal bent by means of a low-amplitude acoustic wave) as 
an undulator for a free electron laser was explored.
The main focus was made on the regime when the undulator radiation is
strongly coupled with the ordinary channeling radiation.
This regime is different from the subject of the present discussion.
The essential role of the large-amplitude regime
of the crystalline undulator was not demonstrated in 
these papers.

\subsection{Quasi-classical description of the crystalline undulator problem}
\label{BaierKatkov}

The important issue of the study of the radiation formed in a 
crystalline undulator concerns the choice of the formalism used to 
describe the phenomenon.
This point could have been regarded as merely a technical one but it is 
not so.
Contrary to the case of conventional undulators, based on the action of
magnetic fields, the physics of crystalline undulators is, basically, 
a newly arisen field of research.
Therefore, any theoretical study of the effect, which pretends to go
a bit farther than purely academic considerations, must contain a great
part of numerical analysis and numerical data on the basis of which real 
experimental investigations can be envisaged.
In turn, to obtain the reliable data it is necessary to choose a 
theoretical tool which allows, on the one hand, to treat adequately all 
principal physical phenomena involved into the problem, and, on the other 
hand, to carry out numerical analysis of the obtained analytical expressions.
In the crystalline undulator problem there are three basic phenomena which
must be accurately described. These are: (i) the motion of an 
ultra-relativistic particle in an external (strong) field,
(ii) the process of photon emission by the particle, 
(iii) the problem of the radiative
recoil, which results in the radiative losses of the projectile.
  
The most rigorous approach to tackle (theoretically) these problems
is the one based on quantum electrodynamics (see, e.g. \cite{Land4}), 
where the amplitude of the process is described in terms of a single 
free-free matrix element of the photon emission taken between the 
initial and final states of an ultra-relativistic particle in the 
interplanar field. 
The main (technical) limitation of this approach appear due to the fact that 
in the ultra-relativistic limit, when $\gamma \gg 1$ the number of the 
energy levels related to the transverse motion in the effective potential 
increases significantly.
Consequently, an accurate description, i.e. numerical calculations,  
of the particle dynamics becomes a formidable task. 
It is exactly this sort of difficulties which resulted in the absence
of {\it any} numerical analysis and data for the emission spectra 
in the papers \cite{BaryshevskyDubovskayaGrubich1980,IkeziLinLiuOhkawa1984,
Dedkov94},
where the radiation formed in the crystalline undulator was treated 
in terms of quantum electrodynamics.

Another option is to study the problem within 
the framework of classical electrodynamics (see, e.g. \cite{Landau2}).
This method was used in the early works 
\cite{KaplinPlotnikovVorobev1980,BogaczKetterson1986}, and 
was later applied in 
\cite{AvakianGevorgyanIspirianIspirian1998,
AvakianGevorgianIspirianIspirian2001,
KorhmazyanKorhmazyanBabadjanyan2004}.
In connection to the crystalline undulator problem the purely
classical description is valid if
(i) the characteristic energy of the projectile in an external field,
$\hbar \tilde{\omega}$, is much less than its total energy, 
$\E = m \gamma c^2$,
and, (ii) the radiative recoil, i.e. the change of the projectile
energy due to the photon emission, is neglected.
The first condition is well fulfilled in the case of the ultra-relativistic 
particle channeling in a crystal.
Indeed, typical values of $\hbar \tilde{\omega}$ are equal, 
in the order of magnitude, to the depth of the interplanar potential well. 
The latter varies from several eV, for the crystals of made of light elements
(e.q. LiH crystal, see \cite{LiH,KorhmazyanKorhmazyanBabadjanyan2003}), 
up to $10^2$ eV for heavy crystals like W
(see, e.g., \cite{BiryukovChesnokovKotovBook}). 
Therefore, $\hbar \tilde{\omega}/\varepsilon \ll \gamma^{-1} \ll 1$.
The role of the radiative recoil is described by the ratio
$\hbar \omega /\varepsilon$.
Purely classical description implies that $\hbar \omega \rightarrow 0$.
Although practical implementation of the classical treatment is comparatively 
simple, in application to the crystalline undulator it is fully approved 
in the so-called dipole-limit (see e.g. \cite{Kniga,Baier}), 
when the undulator parameter is small, $p<1$, and 
all the undulator emission occurs in the fundamental harmonic.
Such an assumption leads to a considerable narrowing of the parameters
of the crystalline radiation and, also, disregards the possibility to generate
the emission in higher harmonics. 
Another disadvantage of the approach based on classical mechanics and
electrodynamics is that it fails when the energy of the projectile becomes
sufficiently large. For a positron this means $\E \leq 10$ GeV.
For this energies the probability of the emission of the 
gamma-quanta of energy $\hbar \omega \leq \varepsilon$ via the 
mechanism of the channeling radiation cannot be neglected.

The third approach, which can be used in studying of the radiative processes
occuring in external fields in ultra-relativistic domain, 
was developed by Baier and Katkov in the late 1960s' \cite{Baier67},
and was called by the authors `the operator quasi-classical method'. 
The details of this formalism can be found also in \cite{Baier,Land4}.

>From the practical viewpoint, the advantage of the quasi-classical method 
is that it justifies the classical description of the motion
of an ultra-relativistic particle in an external field (i.e., the use 
of the trajectories rather than the wavefunctions), and,
simultaneously, takes into account the effect of the radiative
recoil. 
Thus, the quasi-classical approach neglects the the terms 
$\hbar \tilde{\omega}/\varepsilon$, 
but it explicitly takes into account the quantum
corrections due to the radiative recoil in the whole range of the
emitted photon energies, except for the
extreme high energy tail of the spectrum.
Using this method the spectra of photons and electron-positron pairs
in linear crystals were successfully described \cite{Baier}.
It was also applied to the problem of a synchrotron-type radiation
emitted by an ultra-relativistic projectile channeling in
a non-periodically bent crystal \cite{Arutyunov,SolovyovSchaferGreiner}.

In \cite{Korol98}-\cite{Darmstadt01}
this general formalism of Baier and Katkov was used for theoretical and
numerical description of the spectral and angular distribution of the 
crystalline undulator radiation, total photon emission spectra and the 
radiative energy losses of positrons channeling through the
periodically bent crystal.

\section{Crystalline undulator radiation}
\label{CrystUndRad}

To illustrate the crystalline undulator radiation phenomenon, 
let us consider the spectra of spontaneous radiation emitted during
the passage of positrons through periodically bent crystals.
The results presented below clearly demonstrate
the validity of the statements made in 
\cite{Korol98,Korol99,EnLoss00,TotSpect00,Dechan01} 
and summarized in section \ref{AllConditions} above, 
that the properties of the undulator radiation 
can be investigated separately from 
the ordinary channeling radiation.

\begin{figure}
\begin{center}
\epsfig{file=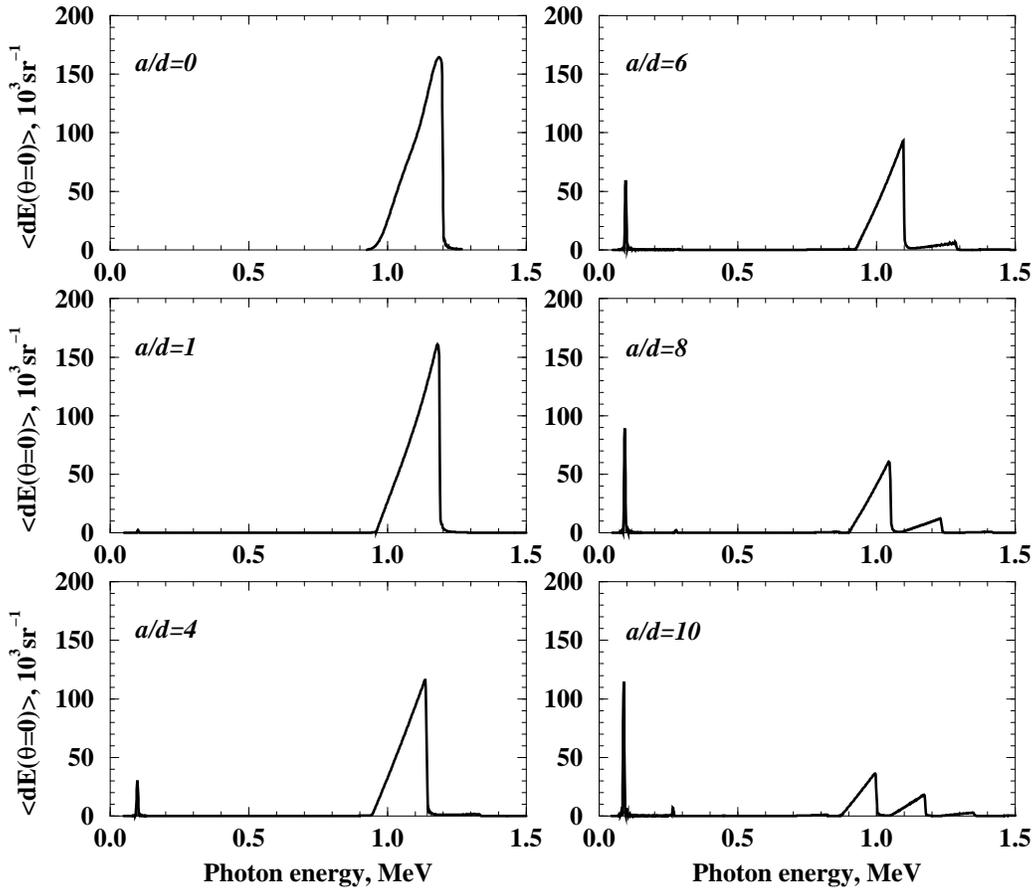 ,width=13cm}
\caption{
Spectral distribution 
of the total radiation emitted in the forward
direction ($\vartheta=0^{\circ}$) 
for $\E=0.5$ GeV ($\gamma \approx 10^3$) positron 
channeling in Si along the (110) crystallographic planes 
calculated at different $a/d$ ratios.
Other parameters are given in the text.
the crystal length is $L=3.5\times 10^{-2}$ cm.}
\label{figure8_jpg}
\end{center}
\end{figure}
The calculated spectra of the radiation emitted in the forward
direction (with respect to the $z$-axis, see figure \ref{figure1_jpg})
in  the case  of $\E=0.5$ GeV planar channeling 
in Si along (110) crystallographic planes 
and for the  photon energies from 45 keV to 1.5 MeV are presented in 
figures \ref{figure8_jpg} \cite{TotSpect00}.
The ratio $a/d$ was varied within the interval $a/d = 0\dots 10$
(the interplanar spacing is 1.92 \AA). 
The case $a/d=0$ corresponds to the straight channel.
The period $\lambda$ used for these calculations
equals to $2.33\times 10^{-3}$ cm. 
The number of undulator periods and crystal length 
were fixed at $N=15$ and $L=N\, \lambda = 3.5\times 10^{-2}$ cm. 
These data are in accordance with the values allowed by 
(\ref{AllConditions.1}) (see also figure \ref{figure5_jpg}).

The spectra correspond to the total radiation, which accounts
for the two mechanisms, the undulator and the channeling.
They were calculated using the quasi-classical method \cite{Baier,Baier67}.
Briefly, to evaluate the spectral distribution the following
 procedure was adopted 
(for more details see \cite{TotSpect00,Durham00,Darmstadt01}).
First, for each $a/d$ value  the spectrum was calculated for 
individual trajectories of the particles. 
These were obtained by solving the relativistic equations of motion
with both the interplanar and the centrifugal potentials taken into
account.
We considered two frequently used \cite{Gemmell}
analytic forms for the continuum interplanar potential, 
the harmonic and the  Moli\`ere potentials calculated at 
the temperature $T=150$ K to
account for the thermal vibrations of the lattice atoms.
The resulting radiation spectra were obtained by averaging over all
trajectories. 
Figures \ref{figure8_jpg} correspond to the spectra obtained
by using the Moli\`ere approximation for interplanar potential.

The first graph in figure \ref{figure8_jpg} corresponds to the case of 
zero amplitude of the bending (the ratio $a/d=0$) and, hence, 
presents the spectral
dependence of the ordinary channeling radiation only. 
The asymmetric shape of the calculated channeling radiation 
peak, which is due to the strong anharmonic character of the Moli\`ere
potential, bears close resemblance with the experimentally measured spectra
\cite{Uggerhoj1993}.
The spectrum starts at $\hbar\omega\approx 960$ keV, reaches its
maximum value at $1190$ keV, and steeply cuts off at  $1200$ keV. 
This peak corresponds to the radiation into the first harmonic of 
the ordinary channeling  radiation (see e.g. \cite{Kumakhov2}), and 
there is almost no radiation into higher harmonics. 

Increasing the $a/d$ ratio leads to the modifications in the 
radiation spectrum. 
The changes which occur are:
(i) the lowering of the channeling radiation peak,
(ii) the gradual increase of the intensity of undulator radiation due to the
crystal bending.

The decrease in the intensity of the channeling radiation
is related to the fact that the increase of the amplitude $a$ of the bending
leads to lowering of the allowed maximum value of the channeling 
oscillations amplitude $a_c$
(this is measured with respect to the centerline  of the bent
channel) \cite{EnLoss00,BiryukovChesnokovKotovBook}.
Hence, the more the channel is bent, the lower the allowed values
of $a_c$ are, and, consequently, the less 
intensive is the channeling radiation, which is
proportional to $a_c^2$ \cite{Baier}.

The undulator radiation related to the motion of the
particle along the centerline of the periodically bent channel is absent
in the case of the straight channel (the graph $a/d=0$), and is almost 
invisible for comparatively small amplitudes (see the 
graph for $a/d=1$). 
Its intensity, which is proportional to $(a/d)^2$,
gradually increases with the amplitude $a$. 
For large $a$ values ($a/d \sim 10$) 
the intensity of the first harmonic of the undulator radiation
becomes larger than that of the channeling radiation.
The undulator peak is located at much lower energies,
$\hbar\omega^{(1)} \approx 90$ keV, and has the width 
$\hbar\Delta \omega \approx 6$  keV which is almost 40 times less than 
the width of the peak of the channeling radiation. 

\begin{figure}
\begin{center}
\epsfig{file=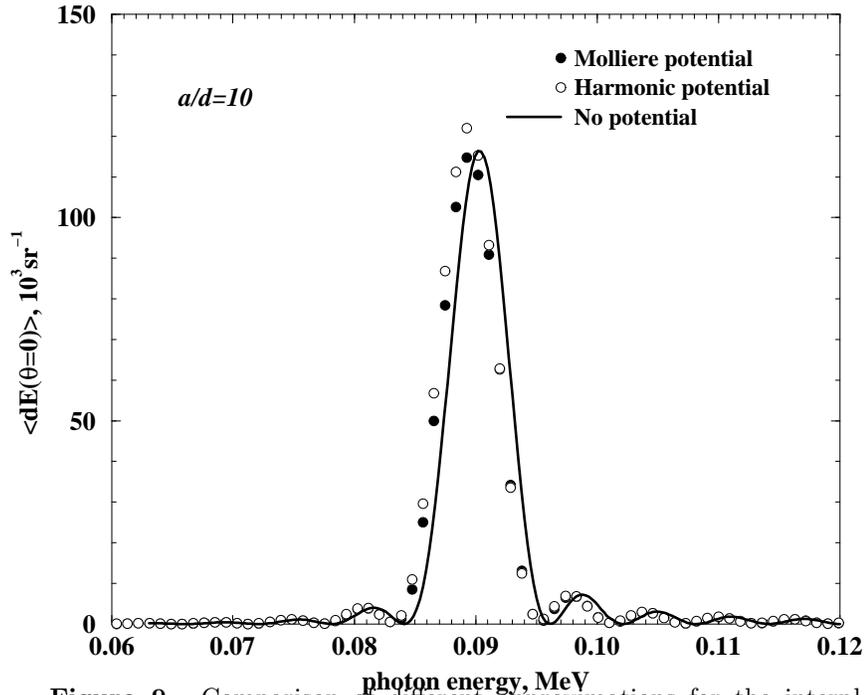 ,width=13cm}
\vspace*{-1.5cm}
\caption{
Comparison of different approximations for the interplanar potentials
used to calculate the total radiative spectrum in vicinity of the 
first harmonic of the undulator radiation.
The ratio $a/d=10$, other parameters as in figure \protect\ref{figure8_jpg}.
}
\label{figure9_jpg}
\end{center}
\end{figure}
It is important to note that the position of sharp undulator radiation 
peaks, their narrow widths, and the radiated intensity are, practically, 
insensitive to the choice of the approximation used 
to describe the interplanar potential. 
In addition,  provided the first two conditions from (\ref{AllConditions.1})
are fulfilled,
these peaks are well separated (in the photon energy scale) from 
the peaks of the channeling radiation. 
Therefore, if one is only interested in the spectral distribution of the 
undulator radiation, one may disregard the channeling oscillations 
and to assume that the projectile moves along the centerline of 
the bent channel \cite{Korol98,Korol99}.
This statement is illustrated by \ref{figure9_jpg} \cite{TotSpect00}
where we compare
the results of different calculations of the radiative spectrum 
in vicinity of the first harmonic of the undulator radiation 
in the case $a/d=10$.
All parameters are the same as in figure \ref{figure8_jpg}.
The filled and open circles represent the results of evaluation of 
the total spectrum of radiation 
accompanied by numerical solution of the equations of motion for
the projectile within the Moli\`ere (filled circles) and the harmonic 
(open circles) approximations for the interplanar potential.
The solid line corresponds to the undulator radiation only.
For the calculation of the latter it was assumed that the trajectory of 
a positron,  $y(z) = a\sin(2\pi\lambda/z)$,
coincides with the centerline of the bent channel 
(see figure \ref{figure1_jpg}).
It is clearly seen that the more sophisticated treatment has 
almost no effect on the profile of the peak obtained by 
means of simple formulae describing purely undulator 
radiation \cite{Korol98,Korol99}. 
Moreover, the minor changes in the position and the height of the
peak can be easily accounted for by introducing the effective undulator
parameter  and (in the case of the harmonic 
approximation) the effective undulator amplitude \cite{EnLoss00}.

\begin{figure}
\begin{center}
\epsfig{file=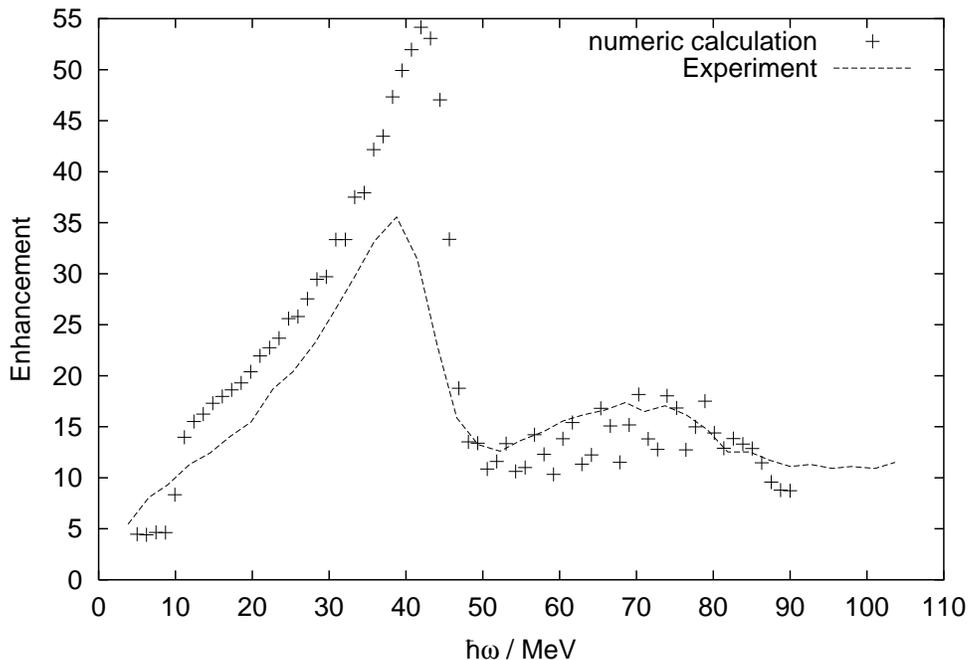 ,width=13cm}
\caption{
Comparison of the experimentally  measured spectrum
\protect\cite{BakEtal1985,Uggerhoj1993} 
and the results of the calculation \protect\cite{Durham00,Darmstadt01}
for 6.7 GeV positrons in Si(110).}
\label{figure10_jpg}
\end{center}
\end{figure}
To check the numerical method, which was developed in
\cite{TotSpect00} for the calculation of the total emission spectrum 
of ultra-relativistic positrons in a crystalline undulator,
we calculated the spectrum of the channeling radiation for 6.7 GeV
positrons in Si(110) integrated over the emission angles. 
Figure \ref{figure10_jpg} shows the experimental data
\cite{BakEtal1985,Uggerhoj1993} and the results of our calculations
\cite{Durham00,Darmstadt01}
normalized to the experimental data at the right wing of the spectrum.
The height of the first harmonic is overestimated in our
calculations. 
The calculations performed in \cite{BakEtal1985} gave a similar result. 
This disagreement arises likely due to the neglect
of multiple collisions which were accounted for 
neither in \cite{Durham00,Darmstadt01} nor in \cite{BakEtal1985}. 
However, the shape and the location of the first harmonic of the channeling
radiation are described quite well. 

\begin{figure}
\begin{center}
\epsfig{file=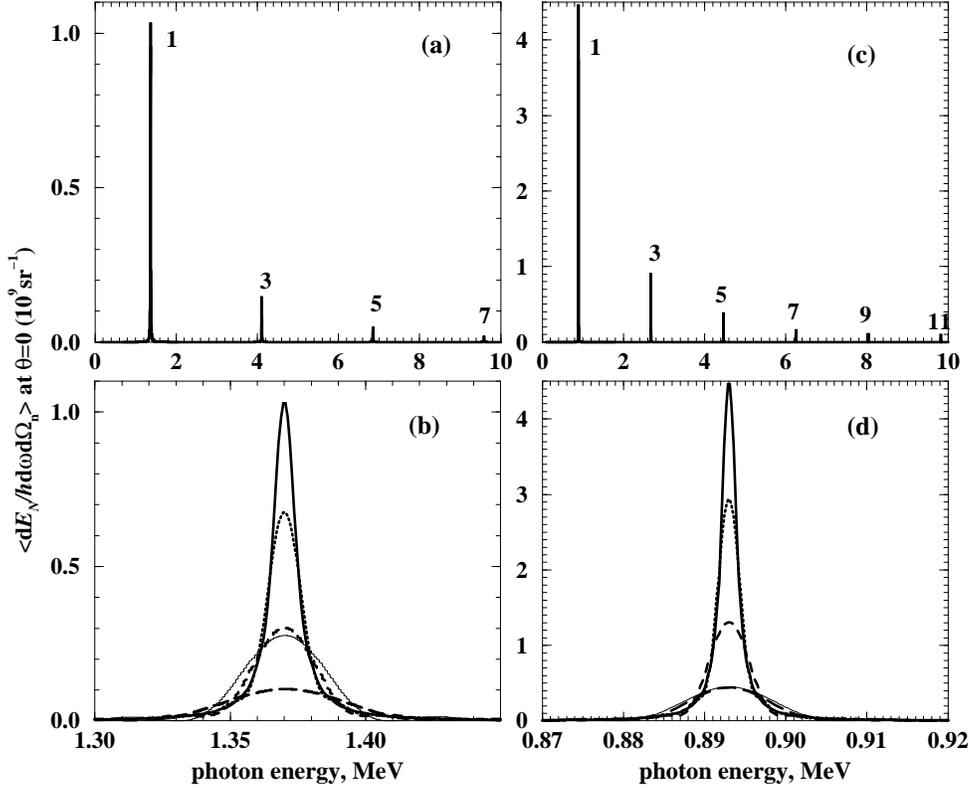 ,width=13cm}
\caption{Spectral distribution (in $10^9\,{\rm sr}^{-1}$)
of the undulator radiation at $\vartheta=0$ 
for 5 GeV positron channeling along periodically bent $(110)$ planes in 
Si (figures (a) and (b)) and W (figures (c) and (d)) crystals.
The $a/d$ ratio is equal to 10.
Other parameters used are presented in table \protect\ref{Table2}.
The upper figures (a) and (c) reproduce 
$\left\langle {\d E_N/\hbar\, \d\omega\,\d\Omega_{\bf n}}\right\rangle $
in the wide ranges of $\omega$ and correspond to $N=4\, N_d$.
The numbers enumerate the harmonics (in the case of the forward
emission the radiation occurs only in odd harmonics).
The profiles of the first harmonic peak (figures (b) and (d)) are
plotted for 
$N=4\, N_d$ (solid lines),
$N=2\, N_d$ (dotted lines),
$N=N_d$ (dashed lines),
$N=N_d/2$ (long-dashed lines). 
}
\label{figure11_jpg}
\end{center}
\end{figure}

The intensity and the profile of the peaks of the undulator radiation 
are defined,  to a great extent, by the magnitude of
the dechanneling length.
In \cite{Dechan01} a more sophisticated, than in \cite{Korol98,Korol99},
theoretical  and numerical analysis of this influence was presented.
In particular, we solved the following problems:
(a) simple analytic expression was evaluated for spectral-angular 
distribution of the undulator radiation which contains, as a parameter, 
the dechanneling length $L_d$,
(b) the simulation procedure of the dechanneling process of a positron 
in periodically bent crystals was presented,
(c) the dechanneling lengths were calculated for 5 GeV positrons 
channeling in Si, Ge and W crystals along the periodically bent 
crystallographic planes,
(d) the spectral-angular  and spectral distributions of the undulator
radiation formed in crystalline undulator were calculated in a broad range 
of the photon energies and for various $a$, $\lambda$ and $C$.

\begin{table}
\caption{The values of the parameter $C$ (see (\ref{Condition.1})),
undulator period $\lambda$,
the dechanneling length $L_d(C)$,
the number of undulator periods $N_d=L_d(C)/\lambda$ within $L_d(C)$,
the undulator parameter $p$ (see (\ref{Om0andP})),
and the fundamental harmonic energy (see (\ref{harmonics.1} with
$\vartheta=0^{\circ}$))
used for the calculation \cite{Dechan01} of the spectra
presented in figures \ref{figure11_jpg}(a)-(d).}
\begin{indented}
\item[]\begin{tabular}{@{}rrrrrrrrr}
\br
Crystal&  $C$&$\lambda$ & $L_d(C)$&$N_d$&$p$ & $\hbar\om_1$ \\
       &     & $\mu$m   &  cm     &     &    &  MeV    \\
\br
 Si    & 0.15& 63.0     & 0.321   &51   & 1.87&1.37 \\
  W    & 0.05& 42.2     & 0.637   &151  & 3.26&0.89 \\
\br
\end{tabular}
\end{indented}
\label{Table2}
\end{table}

To illustrate the results obtained in \cite{Dechan01}, in figures 
\ref{figure11_jpg}(a)-(d) we present the spectral distribution of the 
undulator radiation emitted along the undulator axis,
$\hbar^{-1}\langle \d E_N/\d\om\,\d\Om_{\bf n}\rangle_{\vartheta=0^{\circ}}$, 
for 5 GeV positron channeling along (110) planes in Si and W crystals.
The spectra correspond to the ratio $a/d$, where
$d=1.92$ \AA \, for Si and $d=2.45$ \AA\, for W.
The values of other parameters, used in the calculations,
are given in table \ref{Table2}.
The values of the dechanneling lengths, $L_d(C)$, were obtained
in \cite{Dechan01} by means of
the simulation procedure of the dechanneling process of a positron 
in periodically bent crystals.

The upper figures, \ref{figure11_jpg}(a) and (c),
illustrate the spectral distributions in Si and W over a wide 
range of emitted photon energy, and corresponds 
to the crystal length, $L$, exceeding the dechanneling length by
a factor of 4: $L=4L_d(C)$.
Each peak corresponds to emission into the odd harmonics,
the energies of which follow from
the relation $\om_k=k\, \om_1$, $k=1,3,\dots$.
The difference in the magnitudes of the undulator parameters
for Si and W (see table \ref{Table2}) explains  number of the harmonics
visible in the spectra.
It is seen that all harmonics are well separated: the distance
$2 \hbar\omega_1$  between two neighbouring peaks is $2.74$ MeV for Si and 
$1.78$ MeV in the case of W,
whilst the width of each peak $\hbar \Delta\omega$
is $\approx 8.7$ keV for Si and  $\approx 2.5$ keV for W.

Figures \ref{figure11_jpg}(b) and (d) exhibit, in more detail, 
the structure of the first-harmonic peaks. 
For the sake of comparison we plotted the curves corresponding to
different values of the undulator periods. 
It is seen that for $N>N_d$ the intensity of the peaks is no longer
proportional to $N^2$, as it is in the case of the ideal undulator without
the dechanneling of the particles \cite{Kniga}.
For both Si and W crystals, the intensities of the radiation
calculated at $N\longrightarrow \infty$ exceed those at 
$N=4\,N_d$ (the thick full curves in the figures) only by several per cent. 
Thus, the full curves correspond to almost saturated intensities which 
are the maximal ones for the crystals used, projectile energies and
the parameters of the crystalline undulator.
For a more detailed discussion see paper \cite{Dechan01}.

\section{Stimulated emission from a crystalline undulator}
\label{StimulatedEmission}

As demonstrated in \cite{Korol98,Korol99},
the scheme illustrated by figure \ref{figure1_jpg} allows to consider
a possibility to generate stimulated emission of high energy
photons by means of a bunch of ultra-relativistic positrons moving in
a periodically bent channel. 
The photons, emitted in the forward
direction ($\vartheta=0$) at the points of the maximum curvature of the bent
channel, travel parallel to the beam and, thus, stimulate the 
photon generation in the vicinity of all successive maxima and minima. 
This mechanism of the radiation stimulation is 
similar to that known  for a free-electron laser  
(see, e.g. \cite{SaldinSchneidmillerYurkov1995}), 
in which the periodicity of a trajectory 
of an  ultra-relativistic projectile 
is achieved by applying a spatially periodic magnetic field. 
Also from the theory of FEL it is known \cite{Madey71}, that the 
stimulation occurs at  the frequencies of the harmonics 
of the spontaneous emission, $\omega_k=k\, \om_1$, $k=1,2,\dots$.
The frequency of fundamental harmonic, $\om_1$, is defined in 
(\ref{harmonics.1}).
In \cite{Korol98,Korol99} and, also, in a more recent paper
\cite{KorolSolovyovGreiner2003_2} it was shown, that it is possible 
to separate the stimulated photon emission in the crystalline 
undulator from the ordinary channeling radiation
in the regime of large bending amplitudes $a \gg d$.
This scheme of the stimulated photon emission  allows to generate
high energy photons up to MeV region and, thus,
we call it as a Gamma-laser.
As a further step in developing the ideas proposed in these papers,
the study, carried out in 
\cite{AvakianGevorgianIspirianIspirian2001},
was devoted to the investigation of the influence of the 
beam energy spread on the characteristics of the stimulated emission
in crystalline undulators. 

In the regime of low amplitudes, $a < d$,
the idea of using a periodically bent crystal as an undulator
for a free electron laser was explored in \cite{BogaczKetterson1986}. 
In this regime the intensity of the undulator radiation 
is relatively small compared with the channeling radiation.
However, it is possible to match the undulator frequency to that of 
the channeling motion.
This results in a resonant coupling of the emissions via the two 
mechanisms, which leads to the enhancement of the gain factor.

Let us review the results obtained in 
\cite{Korol98,Korol99,KorolSolovyovGreiner2003_2}.
To do this we first outline the derivation of the general expression for 
the gain factor in an undulator, and, after accounting for the 
conditions (\ref{AllConditions.1}),
estimate gain for the crystalline undulator.
For the sake of simplicity we consider the stimulated emission 
for the fundamental harmonic only, and, also, consider the 
emission in the forward direction. 
In the formulae below, we use the notation $\om$ instead of 
$\om_1$ for the fundamental harmonic frequency.

The gain factor, $g(\omega)$, defines the increase in the total number, 
${\cal N}$,  of the emitted photons at a frequency 
$\omega$ due to stimulated emission by the particles of the beam:
$d {\cal N}=g(\omega)\, {\cal N}\,d z$.
The general expression for the quantity $g(\omega)$ is
\begin{eqnarray}
g(\omega) 
= 
n\,
\left[\sigma_e(\E,\E-\hbar\om) - \sigma_a(\E,\E+\hbar\om)\right],
\label{GeneralExpressionForGain.3}
\end{eqnarray}
where $\sigma_e(\E,\E-\hbar\om)$ and $\sigma_a(\E,\E+\hbar\om)$
are the cross sections of, correspondingly,
the spontaneous emission and absorption of the
photon by a particle of the beam, $n$ stands for
the volume density (measured in cm$^{-3}$) of the beam particles.
By using the known relations between the cross sections
$\sigma_{e,a}$ and the spectral-angular intensity of the emitted 
radiation \cite{Land4}, one derives the following expression for the gain:
\begin{equation}
g =
-(2\pi)^3\, {c^2 \over \omega^2}\, n\,
{d \over d\E}
\left[
{d E \over d \omega\, d \Omega }\right]_{\vartheta=0}
\, \Delta \omega \, \Delta\Omega.
\label{GeneralExpressionForGain.4}
\end{equation}
Here $d E / d \omega\, d \Omega$ is the spectral-angular intensity
of the radiation,
$\Delta \omega$ is the width of the first harmonic peak,
and $\Delta\Omega$ is the effective cone (with respect to the
undulator axis) into which the emission of the $\om$-photon occurs.
Note that expression (\ref{GeneralExpressionForGain.4}) is derived
under the assumption that the photon energy is small compared to the
energy of the particle,  $\hbar\omega \ll \E$.

For an undulator of the length $L$ the total increase in the number of
photons is
\begin{eqnarray} 
\calN
=
\calN_0
\,
{\rm e}^{G(\om)L},
\label{GeneralExpressionForGain.5}
\end{eqnarray} 
where $G(\om)= g(\om) L$ is the total gain on the scale $L$.
The expression for $G(\om)$ follows 
from (\ref{GeneralExpressionForGain.4}) (the details of derivation
one finds in \cite{Korol99,Baier}):
\begin{eqnarray} 
G(\om)
=
n\,
(2\pi)^3r_0\,{Z^2 \over M}\,
{L^3  \over \gamma^3\, \lambda}
\cdot
\left\{
\begin{array}{cc} 
1   &{\rm if}\ p^2>1\\
p^2 &{\rm if}\ p^2<1
\end{array}
\right.
\label{GeneralExpressionForGain.6}
\end{eqnarray} 
where
$r_0=2.8\cdot10^{-13}$ cm is 
the electron classical radius,
$Z$ and $M$ are 
the charge and the mass of a projectile in
the units of elementary charge and electron mass.
Note the strong inverse dependence on $\gamma$ and $M$ which is 
due to the radiative recoil, and the 
proportionality of the gain  
to $L^3$ and to the squared charge of the projectile $Z^2$.
 
The main difference, of a principal character, between a conventional
FEL and a FEL-type device based on a crystalline undulator is that
in the former the bunch of particles and the photon flux both travel
in vacuum whereas in the latter they propagate in a crystalline medium.
Consequently, in a conventional FEL one can, in principle, 
increase infinitely the length of the undulator $L$.
This will result in the  increase of the total gain and the 
number of undulator periods $N$, (\ref{Condition.2}). 
The limitations on the magnitude of $L$ in this case are mainly of a 
technological nature.

The situation is different for a crystalline undulator, 
where the dechanneling effect and the photon attenuation lead to 
the decrease of $n$ and of the photon flux density  
with the penetration length and, therefore,  
result in the limitation of the allowed $L$-values.
The reasonable estimate of $L$ is given by the condition
(\ref{Conditions.5}).
In turn, this condition, together with the estimate
(\ref{Condition.La}), defines the ranges of photon energies 
for which the operation of a crystalline undulator is realistic.
These ranges are:
\begin{itemize}
\item  High-energy photons: $\hbar\omega > 10$ keV when $L_a > 0.01$ cm;
\item  Low-energy photons: $\hbar\omega < I_0\leq 10$ eV. 
\end{itemize}

In the regime of high-energy photons (the gamma-laser regime)
the stimulation of the emission
must occur during a single pass of the bunch of the particles through
the crystal. 
Indeed, for such photon energies there are no mirrors, and, therefore,
the photon flux must develop simultaneously with the bunch propagation.
In the theory of FEL this principle is called 
'Self-Amplified Spontaneous Emission' (SASE)
\cite{SaldinSchneidmillerYurkov1995,Pellegrini1984}
and is usually referred as the FEL operation in the high gain regime,
which implies that $G(\om)>1$ 
to ensure that the exponential factor in 
(\ref{GeneralExpressionForGain.5}) is large. 
In this case the quantity $\calN_0$ denotes the
number of photons which appear due to the spontaneous emission 
at the entrance part of the undulator.

For  $\hbar\omega < I_0\leq 10$ eV there is no 
principal necessity to go beyond 1 for the magnitude of $G(\om)$
during a single pass.
Indeed, for such photons there is a possibility to use mirrors
to reflect the photons.
Therefore, the emitted photons, after leaving the undulator can be 
returned back to the entrance point to be used for further stimulation 
of the emission by the incoming projectiles.

Below we present the results of numerical calculations
of the parameters of the undulator (the first harmonic energy and 
the number of periods) and of the volume density $n$ of the 
bunch particles needed to achieve $G(\om)=1$.
The calculations were performed for relativistic
positrons, muons, protons and heavy ions and took into account all the
conditions summarized in (\ref{AllConditions.1}).
The results presented correspond to the cases 
of the lowest values of $n$ needed to ensure
$G(\om)=1$ and which, simultaneously, produce the largest available 
values of $N$.

\subsection{High-energy photons: the gamma-laser regime}
\label{GammaLaserRegime}

Detailed analysis of the conditions (\ref{AllConditions.1}) and 
demonstrates, that to optimize the parameters of the stimulated
emission in the photon energy range
$\hbar\omega > 10$ keV in the case of a positron channeling in 
a periodically bent crystal one should consider the
following ranges of parameters:
$\gamma=(1\dots 5)\times 10^3$,
$a/d=10\dots20$,
$C=0.1\dots0.3$,
which are common for all the crystals which we have investigated.
For these ranges the energy of the first harmonic
(see (\ref{harmonics.1})) lies within the interval 
$50 \dots 150$ keV, and the length of the 
undulator can be taken equal to the dechanneling length because of the
inequality $L_d(C) < L_a(\om)$, valid for such $\om$.

\begin{figure}
\begin{center}
\epsfig{file=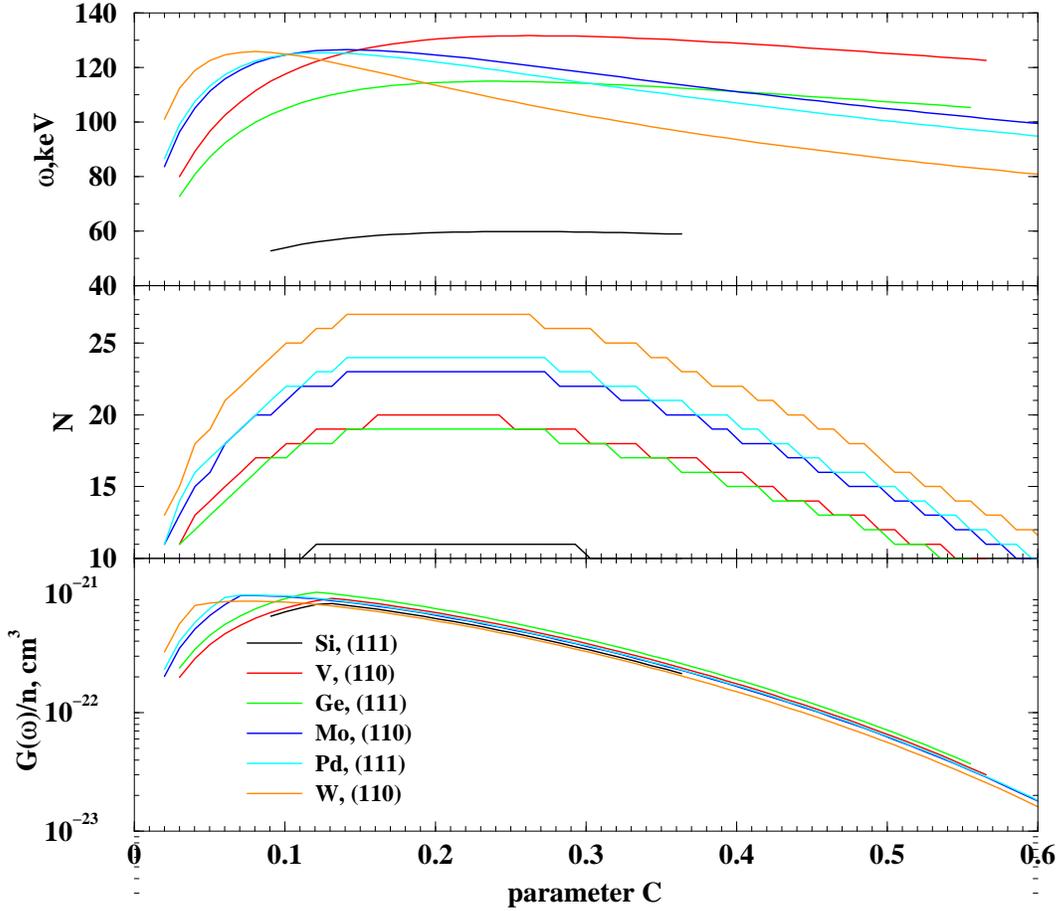, width=14cm}
\vspace*{-0.5cm}
\caption{
First harmonic energy, $\om$,
number of undulator periods, $N$, and the ratio
$G(\om)/n$ in cm$^{3}$ versus $C$ for 0.5 GeV positron channeling
in various channels as indicated.}
\label{figure12_jpg}
\end{center}
\end{figure}

Results of calculations are presented in figure
\ref{figure12_jpg}, 
where the dependences of the first harmonic energy, $\hbar\om$,
the number of undulator periods, $N$, and
the ratio $G(\om)/n$  versus the parameter $C$ 
(see (\ref{Condition.1}))
are presented for various crystals.
The data correspond to the ratio $a/d=20$ except for the case of 
Si for which $a/d=10$.

For each crystal the curves $\hbar\om$ and $G(\om)/n$ were truncated 
at those $C$ values for which the number of undulator periods
becomes less than 10 (see the graph in the middle).
It is seen from the bottom graph, that $G(\om)/n$ is a rapidly varying
function of $C$ (note the log scale of the vertical axis).
For all the channels this function attains its maximum value 
$\approx 10^{-21}$ cm$^3$ at $C\approx 0.1$. 
The maximum value of $G(\om)/n$ defines the magnitude of the 
volume density of a positron bunch needed to achieve total gain 
$G(\om)=1$. 
Then it follows from the graph that to achieve the emission stimulation
within the range $\hbar\om=50 \dots 150$ keV 
on the basis of the SASE mechanism it is necessary to 
reach the value $n=10^{21}$ cm$^{-3}$ for a positron bunch of the energy
of several GeV.

At first glance the idea of using a crystalline undulator based on 
the channeling of heavy positively charged particles looks very 
attractive.
Indeed, as it is seen from 
(\ref{DechannnelingAttenuation.3}), (\ref{DechannnelingAttenuation.4})
and (\ref{Condition.Ld}),
the dechanneling length for a heavy particle is $M/Z \gg 1$ times larger 
than that for a positron with the same value of $\gamma$.
This factor, being cubed  in (\ref{GeneralExpressionForGain.6}),
could lead to a noticeable increase of the total gain
(over-forcing, in the cases of $\mu^{+}$ and $p$, 
the small multiplier $Z^2/M$).

However, as it is seen from (\ref{Condition.1}) 
the allowed undulator period $\lambda$ increases with 
a projectile mass: $\lambda > \lambda_{\min} \propto \sqrt{M}$.
In turn, this results in a decrease in the first harmonic energy 
$\om \approx 4\pi c\,\gamma^2/ \lambda \propto 1/ \sqrt{M}$ 
(in the case of a heavy projectile the undulator 
parameter is small, and the term $p^2$ can be disregarded
when calculating $\om$, see (\ref{harmonics.1})).
For  realistic values of relativistic factor, $\gamma \leq 10^3$,
this results in the following restriction on the photon energy:
\begin{eqnarray*} 
\hbar\om \leq \hbar\om_{\max}
\approx
\left\{
\begin{array}{cl} 
50\ \mbox{keV} & \mbox{for}\ \mu^{+} \\
10\ \mbox{keV} & \mbox{for}\ p \\
<1\ \mbox{keV} & \mbox{for a heavy ion }
\end{array}
\right.
\end{eqnarray*} 
For a proton and an ion the range of photon energies is exactly the one  
where the attenuation is very strong.
Therefore,  the crystal length is defined by a small value of 
$L_a(\om)$, see figure \ref{figure6_jpg}.
In the case of $\mu^{+}$ the upper limit of $\hbar\om$ is higher but, 
nevertheless, it leads to a condition $L_a(\om) \ll L_d(C)$, so that
the crystal length also must be chosen as $L=L_a(\om)$.
Although in such conditions it is possible to construct an undulator 
with sufficiently large number of periods, the total gain factor becomes
very small:
\begin{eqnarray*} 
G(\om) \sim
n\cdot
\left\{
\begin{array}{cl} 
10^{-22} & \mbox{for}\ \mu^{+}\\
10^{-26} & \mbox{for}\ p 
\end{array}
\right.
\end{eqnarray*} 
Therefore, it is not realistic to consider the stimulated emission
from a heavy projectile in the high-energy photon range.

\subsection{Low-energy photons}
\label{GainForPositron}

For $\hbar\om \leq I_0$ the photon attenuation becomes small and the
length of the undulator is defined by the dechanneling length of a
particle. 

\begin{figure}
\begin{center}
\epsfig{file=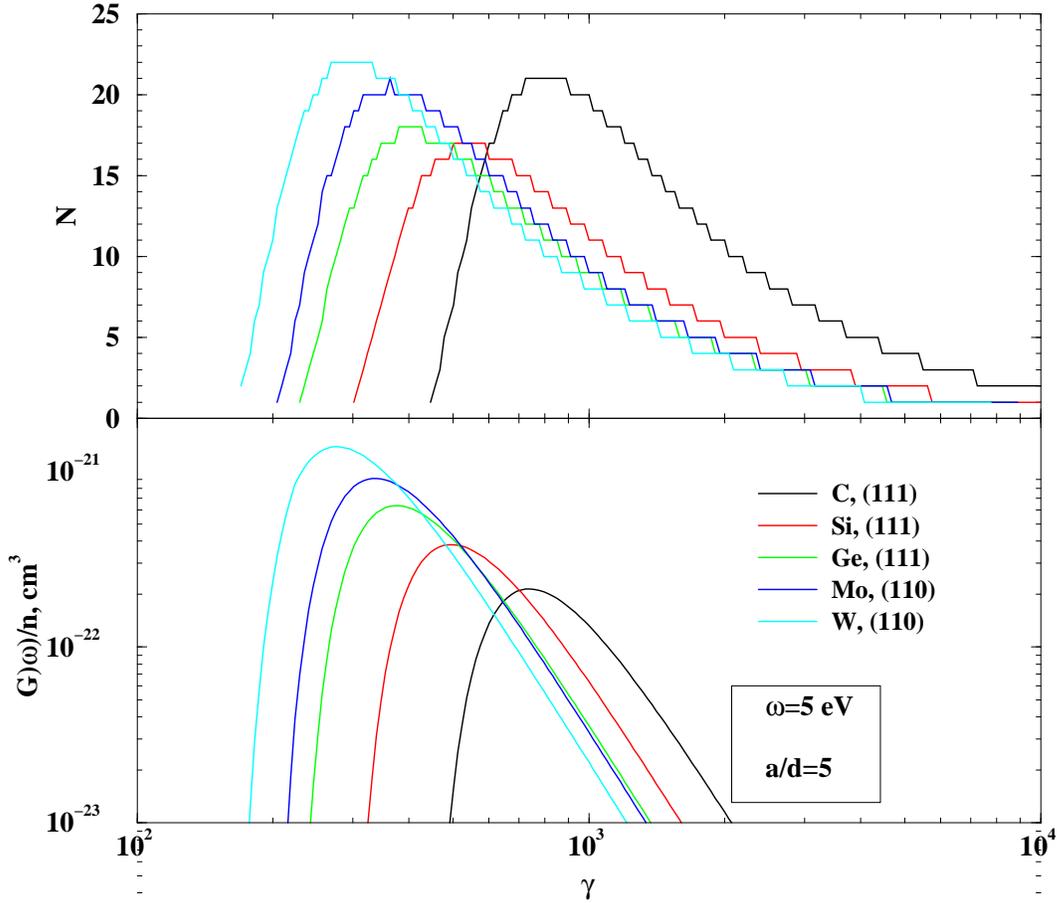, width=14cm}
\vspace*{-1.0cm}
\caption{Values of $N$ and $G(\om)/n$ versus $\gamma$ 
for a positron-based  crystalline undulators in a low-$\om$ region
calculated for various channels as indicated.}
\label{figure13_jpg}
\end{center}
\end{figure}

To illustrate the regime of low-energy stimulated emission
during the positron channeling, 
in figure \ref{figure13_jpg} we present the dependences of 
$N$ and $G(\om)/n$ on the relativistic factor.
The data correspond to a fixed ratio $a/d=5$ and to a fixed energy
of  the first harmonic, $\hbar\om=5$ eV,
which is lower than the atomic ionization potentials
for all crystals indicated in the figures.
The undulator length was chosen to be equal to the 
dechanneling length, 
which is the minimum from  $L_d(C)$ and $L_a(\om)$.
Two graphs in figure \ref{figure13_jpg} demonstrate,
that although the number  of the undulator periods is, approximately, 
independent on the type of
the crystal, the magnitude of the total gain is quite sensitive to
the choice of the channel. 
The highest values of $G(\om)/n$ (and, correspondingly, the lowest
densities $n$ needed to achieve $G(\om)=1$) 
can be achieved for heavy crystals.

Analysis of (\ref{GeneralExpressionForGain.6}) together 
with (\ref{AllConditions.1})
shows, that, in the case of a heavy projectile, 
to obtain the largest possible values of the total gain
$G(\om)$ during a single pass through a crystal 
the following regime can be considered:
(a) moderate values of the relativistic factor,
$\gamma \sim 10\dots 100$;
(b) $C=0.25$ which turns out to be the optimal value of $C$;
(c) $Z\gg 1$, i.e. the best choice is to use a bunch of heavy ions.

\begin{figure}
\begin{center}
\epsfig{file=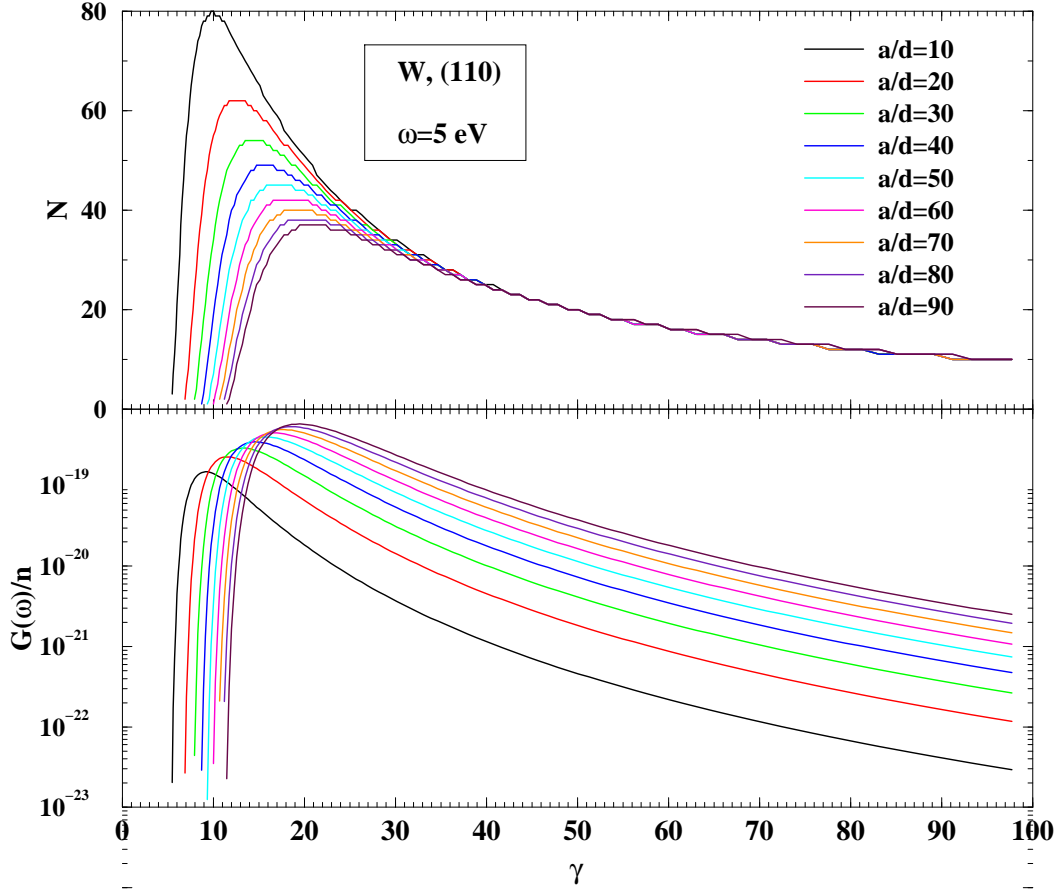, width=14cm}
\vspace*{-0.5cm}
\caption{
Number of periods, $N$, and values of $G(\om)/n$ in cm$^3$ versus $\gamma$
for an ion-based crystalline undulator
in a low-$\om$ region calculated at different $a/d$ values as indicated.
The data are presented for bare $U$ ion channeling along the $(110)$ planes
in a tungsten crystal.}
\label{figure14_jpg}
\end{center}
\end{figure}

In figure \ref{figure14_jpg} we present the dependences of
$N$ and $G(\om)/n$ 
on $\gamma$, lying within the range specified above. 
All curves, which were obtained for different $a/d$ ratios,
refer to the case of $U^{+92}$ channeling in $W$ along the
$(110)$ crystallographic planes.
The energy of the emitted photon is fixed at $5$ eV.
It is seen from the graphs that for all the crystals the most optimal 
range of relativistic factor is $\gamma \sim 10\dots 30$ where both the
number of the undulator periods and the magnitude of $G(\om)/n$ 
noticeably exceed the corresponding values in the case of a positron
channeling, see figure \ref{figure13_jpg}.

Similar analysis, carried out for the case of a proton 
channeling, demonstrates that for the same value of $\gamma$ 
the magnitudes of $G(\om)/n$ are several times higher than those for 
a heavy ion.

Results presented in this section show, that the 
stimulated emission in the low-$\om$ range 
($\hbar\om < I_0 \sim 10$ eV)
can be discussed for all types of positively-charged projectiles.
In this case to achieve the value $G(\om)=1$ within the SASE mode
it is necessary to consider
the densities from $n\approx 5\times10^{18}$  cm$^{-3}$ for heavy ion
beams up to $n\approx 5\times10^{21}$  cm$^{-3}$ for a positron beam.
However, this large numbers can be reduced by orders of magnitudes
if one considers the multi-pass mode of the FEL.
Indeed, there exist the mirrors which allow to reflect photons of the
energy  $\hbar\om < I_0 \sim 10$ eV.
Thus, the emitted photons can be returned back to the entrance 
point and used further to stimulate the emission generated by the particles
of the long bunch. 
The number of the passes, equal approximately to
$L_{bunch}/L$ (here $L_{bunch}$ is the bunch length), can be very large
(up to $10^4$).
Therefore, volume density can be reduced by the factor 
$L/L_{bunch} \ll 1$.

\section{Conclusions}
\label{sec:conclusion}

In this paper we have discussed the feasibility of the
crystalline undulator and Gamma-laser based on it. We have
presented the detailed review covering the development
of all essential aspects of these important ideas.

Firstly, we note that it is absolutely realistic to use a crystalline 
undulator for generating {\it spontaneous} radiation in a wide range
of photon energies.
The parameters of such an undulator, being subject
to the restrictions mentioned in section \ref{Conditions}, can be 
easily tuned by varying the shape function, the energy and the type of 
a projectile and by choosing different channels.
The large range of energies available in modern colliders for various
charged particles, both light and heavy, together with the wide range
of frequencies and bending amplitudes in crystals
allow to generate the crystalline undulator radiation
with the energies from eV up to the MeV region. 

Secondly, it is feasible to obtain {\it stimulated} emission by 
means of a crystalline undulator. 
For a single-pass laser (SASE mode)  high volume densities are needed:
the stimulated emission in the high-$\om$ range ($\hbar\om > 10$ keV) 
demands high volume densities of positrons, 
$n\geq 10^{20}$ cm$^{-3}$.
For this values of $n$ a FEL device based on a crystalline undulator 
can be operated in a single-pass (SASE) mode.

Stimulated emission in the low-$\om$ range 
($\hbar\om < I_0 \sim 10$ eV)
can be discussed for all types of positively-charged projectiles.
In this case 
the large values of the beam densities required 
for the lasing effect 
can be reduced by orders of magnitudes
if one considers the multi-pass mode of the FEL.
Indeed, there exist the mirrors which allow to reflect photons of the
energy  $\hbar\om < I_0 \sim 10$ eV.
Thus, the emitted photons can be returned back to the entrance 
point and used further to stimulate the emission generated by the particles
of the long bunch. 
The number of the passes, equal approximately to
$L_{bunch}/L$ (here $L_{bunch}$ is the bunch length), can be very large
(up to $10^4$).
Therefore, beam density at which the lasing effect arises can be reduced 
by the factor $L/L_{bunch} \ll 1$.

The crystalline undulators discussed in this paper can serve as a
new efficient source for the coherent high energy photon emission. 
As we have pointed out, the present technology is nearly sufficient to
achieve the necessary conditions to construct not only crystalline
undulator, but also the stimulated photon emission source.
The parameters of the crystalline undulator and the Gamma-laser based on it
differ substantially from what is possible to achieve
with the undulators constructed on magnetic fields.

This review clearly demonstrates that experimental effots
are needed for the verification of  numerous theory predications 
outlined in this review. Such efforts will certainly make this field of 
endevour  even more facinating than as it is now and 
possibly will lead to the practical development of
a new type of tunable and monochromatic radiation  sources.

Finally, we mention that not at all theoretical issues for the 
described system  have been solved so far. Thus, the analysys
of dynamics of a high-density positon beam  channeling through 
a periodically bent crystal
in presence of an induced high intensity photon flux
is to be performed in a greater detail. This and many more other
interesting theoretical problems are still open for 
future investigation.



\end{document}